\documentclass[a4paper]{aa}

\usepackage{graphicx}
\usepackage{txfonts}
\usepackage{natbib}
\usepackage{color}
\usepackage{multirow}
\usepackage{stmaryrd}
\usepackage{siunitx}

\newcommand{\invers}[1]{{\sc invers{#1}}}

\newcommand{\llm}{{\sc LLmodels}}

\newcommand{\bz}{$\langle B_\mathrm{z} \rangle$}
\newcommand{\kms}{km\,s$^{-1}$}

\begin{document}

\title{Magnetic field topology and chemical spot \\ distributions of the Ap star HD\,119419%
\thanks{Based on observations collected at the European Southern Observatory, Chile (ESO programs 088.D-0066, 090.D-0256).}}

   \author{N.~Rusomarov
          \and O.~Kochukhov
          \and A.~Lundin}
          
   \institute{Department of Physics and Astronomy, Uppsala University, Box 516, 75120 Uppsala, Sweden \\ \email{oleg.kochukhov@physics.uu.se}}
   \date{Received 8 September 2017; accepted 4 October 2017}

   \titlerunning{Surface structure of the Ap star HD\,119419}
   \authorrunning{N.~Rusomarov et al.}

\abstract
{Analysis of high-resolution spectropolarimetric time-series observations of early-type magnetic stars is currently the most advanced method of obtaining detailed information on their surface magnetic field topologies and horizontal spot distributions.}
{In this study we analyse a new set of high-quality four Stokes parameter observations of the magnetic Ap star HD\,119419 -- a member of the 14~Myr old Lower Cen-Cru association -- for the purpose of studying the surface field topology and mapping the chemical abundance spots.}
{We made use of the circular and linear polarisation data collected for HD\,119419 with the HARPSpol instrument at the ESO 3.6-m telescope. These observations were analysed with a multi-line magnetic diagnostic technique and modelled in detail with a Magnetic Doppler imaging code.}
{We present a new set of high-precision mean longitudinal magnetic field measurements and derive a revised stellar rotational period by comparing our measurements with the literature data. We also redetermine the basic stellar atmospheric parameters. Our four Stokes parameter magnetic inversions reveal a moderately complex surface field topology with a mean field strength of 18~kG and a maximum local strength of 24~kG. A poloidal dipolar component dominates the magnetic energy spectrum of the surface field in HD\,119419. However, significant contributions of the higher-order spherical harmonic components are also present. We show that the dipole plus quadrupole part of the reconstructed field geometry is incapable of reproducing the observed amplitudes and shapes of the Stokes $Q$ and $U$ profiles. The chemical abundance distributions of Fe, Cr, Ti, and Nd, derived self-consistently with the magnetic field geometry, are characterised by large abundance gradients and a lack of clear correlation with the magnetic field structure.}
{This full Stokes vector analysis of HD\,119419 extends the modern hot-star magnetic mapping investigations to an open cluster Ap star with a well-determined age. Further Magnetic Doppler imaging studies of cluster members will allow us to study the field topologies and chemical abundance spots as a function of stellar age.}
   

   \maketitle

\section{Introduction}\label{sec:intro}

Magnetic chemically peculiar (ApBp) stars are the upper main sequence objects with large surface abundance anomalies and strong, stable, globally organised magnetic fields. These fields, thought to have a fossil origin \citep{Braithwaite2004}, have strengths in the range from a few hundred Gauss to several tens of kG. Owing to a coherent, typically quasi-sinusoidal  rotational phase variation of the mean line of sight field component (longitudinal field), magnetic topologies of Ap stars were initially described as oblique dipoles \citep{Stibbs1950}. This approach is still popular with recent studies of large stellar samples \citep{Auriere07p1053,Hubrig2007}. Nevertheless, as observational photo- and spectropolarimetric techniques have matured, enabling more diverse and precise observational constraints, it has become necessary to adopt more complicated dipole plus quadrupole superpositions to properly fit the circular \citep{Landstreet2000p213,Bagnulo2002p1023} and linear \citep{Bagnulo2000p929,Bagnulo2001p889} polarisation measurements of Ap stars.

The ultimate method of modelling both the surface magnetic field geometry and associated chemical element inhomogeneities is to extract information directly from the high-resolution line profiles recorded in several Stokes parameters. Initial attempts at applying this magnetic inversion approach to Ap stars date back to the studies by \citet{Piskunov1983,Piskunov1984} and \citet{Glagolevskii1985}. Subsequently, \citet{Piskunov2002p736} and \citet{Kochukhov2002p868} presented a modern formulation of the Magnetic Doppler imaging (MDI) problem and tested the corresponding inversion code with simulated Stokes parameter data. Their MDI method was then applied to the Stokes $IV$ \citep{Kochukhov2002,Kochukhov2014p83,Kochukhov2015p79,Lueftinger10p71,Oksala2015} and Stokes $IQUV$ \citep{Kochukhov2004p613,Kochukhov10p13,Silvester2014p182,Silvester2015p2163,Rusomarov2015p123,Rusomarov2016p138} observations of several magnetic Ap stars. 

Arguably, one of the most notable outcomes of these MDI investigations was a detection of significantly smaller-scale and higher-contrast local magnetic field structures, typically superimposed on a global dipolar background, than suggested by earlier low-order multipolar fitting studies. However, the number of Ap stars analysed with full Stokes vector spectra, which are essential for unveiling the full extent of the local field complexity, is still very small. This prevents a meaningful analysis of the global and local magnetic field characteristics as a function of stellar mass, age, and rotation. It is therefore imperative to increase the sample of Ap stars investigated with the MDI technique with particular emphasis on the objects with accurately known mass and age, such as ApBp stars in open clusters.

The star HD\,119419 (HR\,5158, HIP\,67036, V827 Cen) is a bright, southern, magnetic ApSiCrEu object and a member of the Lower Cen-Cru (Sco OB2) association with a well-established age of 14~Myr \citep{Landstreet2007}. This star was recognised as a chemically peculiar object by \citet{Jaschek1959} and repeatedly studied for photometric variability in the optical \citep[e.g.][]{Mathys1985} and near-infrared \citep{Catalano1998}. The $\approx$\,2.6~d rotational period of HD\,119419 was first established by \citet{Manfroid1980} and subsequently revised in a number of studies \citep[see][and references therein]{Catalano1998b}.

A strong magnetic field was detected in HD\,119419 by \citet{Thompson1987p219}. Additional longitudinal field measurements were obtained by \citet{Bohlender1993p355}, \citet{Mathys1994p547} and \citet{Mathys97p475}. These studies have established that the reversing longitudinal field of this star exhibits variations from about $-3$~kG to nearly $+2$~kG, indicating a strong surface magnetic field. \citet{Mathys1995p746} reported $\sim$\,20~kG mean quadratic magnetic field. These measurements were revised downwards by \citet{Mathys2017}.

Based on the observed phase curves of the mean longitudinal magnetic field, quadratic field and crossover \citet{Bagnulo1999p158} developed a detailed, non-axisymmetric multipolar field geometry model for HD\,119419, comprising a 17~kG dipole component superposed to a 48~kG quadrupole. This quadrupole-dominated, relatively complex surface field topology was subsequently revised by \citet{Bagnulo2002p1023}. According to their study, the quadrupole to dipole polar field strength ratio, $B_{\rm q}/B_{\rm d}$, is 2.5, the mean surface field strength is about 15~kG and the maximum local field modulus is 40~kG. On the other hand, \citet{Glagolevskii2001} ruled out a purely dipolar field geometry and suggested an off-centred dipolar field configuration from similar modelling of the mean longitudinal and quadratic field variation of HD\,119419. His magnetic field geometry implied an inclination angle close to 90\degr, a small magnetic obliquity and a polar field strength of 39~kG.

With its moderately rapid rotation, strong surface magnetic field, and intriguing evidence of the field complexity, HD\,119419 is an ideal target for MDI in four Stokes parameters. In this paper we present magnetic and surface spot mapping of this star based on a new high-resolution, full Stokes vector spectropolarimetric data set. The rest of the paper is organised as follows. Section~\ref{sec:obs} describes observations and data reduction. Stellar parameters and mean element abundances are discussed in Sect.~\ref{sec:fund}. An analysis of the mean polarisation profiles and revision of the stellar rotational period is presented in Sect.~\ref{sec:lsd}. This is followed by Sect.~\ref{sec:mdi} which describes the methodology and results of the Magnetic Doppler imaging analysis. The paper ends with a summary and conclusions in Sect.~\ref{sec:summary}.

\section{Observations and data reduction}\label{sec:obs}

Spectropolarimetric observations of HD\,119419 were obtained during two observing runs carried out in March-April 2012 and February 2013 at the ESO 3.6\,m telescope with the HARPS spectrograph \citep{Mayor2003p20} and its polarimetric unit HARPSpol \citep{Piskunov11p7,Snik2011p237}. The HARPSpol spectropolarimeter consists of two independent units that measure the circular and linear polarisation respectively, and allow observations in all four Stokes parameters. In total, we acquired 36 individual Stokes parameter observations, which were spread quasi-equidistantly over the entire rotational period of the star, resulting in a good phase coverage with 12 distinct rotational phases.

Each single spectropolarimetric observation of a Stokes parameter consisted of four sub-exposures obtained at different orientations of the quarter- or half-wave plate of the circular and linear polarimeters, respectively. From one such sequence of observations, we derived a single intensity spectrum (Stokes~$I$) and Stokes~$V$, $Q$, or $U$ parameter using the ``ratio'' spectropolarimetric demodulation technique \citep{Bagnulo2009p993}. The diagnostic null polarisation spectrum was derived as well and found not to contain any artefacts. For the subsequent analysis the Stokes $I$ spectra obtained with each polarisation Stokes parameter were averaged together and the phase difference (corresponding to less than 2\% of rotation period) within each full four Stokes parameter observation was ignored.

\begin{table*}
\caption{Journal of spectropolarimetric observations of HD\,119419.}
\begin{tabular}{lccccccccc}
\hline
\hline
\multirow{2}{*}{UT Date} & \multicolumn{3}{c}{HJD\,(2\,455\,000+)} & \multirow{2}{*}{$\overline{\varphi}$}& \multirow{2}{*}{$\delta\varphi$}& \multicolumn{3}{c}{Median S/N}\\
 & $Q$ & $U$ & $V$ &  &  & $Q$ & $U$ & $V$\\
\hline
28-03-2012 & 1014.6646 & 1014.6940 & 1014.6345 & 0.813 & 0.014 &  238 & 238 &240 \\
29-03-2012 & 1015.6428 & 1015.6676 & 1015.6141 & 0.188 & 0.013 &  241 & 225 &262 \\
30-03-2012 & 1016.6629 & 1016.6922 & 1016.6307 & 0.581 & 0.015 &  188 & 187 &172 \\
31-03-2012 & 1017.6248 & 1017.6541 & 1017.5944 & 0.951 & 0.014 &  217 & 207 &195 \\
01-04-2012  & 1018.6407 & 1018.6701 & 1018.6095 & 0.341 & 0.015 &  184 & 202 &200 \\
02-04-2012  & 1019.6613 & 1019.6999 & 1019.6244 & 0.734 & 0.018 &  254 & 242 &228 \\
22-02-2013 & 1345.7403 & 1345.7789 & 1345.7042 & 0.121 & 0.018 &  170 & 168 &156 \\
23-02-2013 & 1346.7444 & 1346.7830 & 1346.7097 & 0.507 & 0.018 &  104 & 224 &126 \\
24-02-2013 & 1347.7391 & 1347.7777 & 1347.7024 & 0.889 & 0.018 &  204 & 209 &202 \\
25-02-2013 & 1348.7374 & 1348.7760 & 1348.7021 & 0.273 & 0.018 &  230 & 228 &196 \\
26-02-2013 & 1349.7433 & 1349.7819 & 1349.7079 & 0.660 & 0.018 &  214 & 242 &180 \\
27-02-2013 & 1350.7321 & 1350.7707 & 1350.6960 & 0.040 & 0.018 &  251 & 210 &230 \\
\hline
\end{tabular}
\label{tab:journal}
\tablefoot{First column gives the UT date at the beginning of each observing night. Heliocentric Julian Dates (HJD) at the mid-exposure are given in columns 2--4 for each observed Stokes parameter. Mean phase, $\overline{\varphi}$, and the maximum difference, $\delta\varphi$, between $\overline{\varphi}$ and phases of individual Stokes parameter observations are presented in columns 5--6. Rotational phases were calculated according to our improved ephemeris (Sect.~\ref{sec:period}). The median S/N for each individual Stokes parameter observation are presented in columns 7--9. The median S/N was calculated using several orders around $\lambda=5500$\,\AA\ where maximum counts were reached.}
\end{table*}

The resulting spectra have resolving power $\lambda / \Delta \lambda$ of approximately 110\,000 and cover the 3780--6910\,\AA\ wavelength region with an 80\,\AA\ gap at 5290\,\AA. Each polarimetric sub-exposure had a length of 600\,s in 2012. For the data obtained in 2013 run, the exposure time varied between 600 and 800\,s to compensate for bad seeing. The representative median signal-to-noise ratio (S/N) measured at 5500\,\AA\ is between 200--250. For two nights (22 and 23 February 2013) the S/N is less than the average value due to the bad seeing that we experienced during those nights. The detailed journal of observations is provided in Table.~\ref{tab:journal}.

The observations of HD\,119419 analysed in this paper were processed with the same set of reduction tools and algorithms as described by \citet{Rusomarov2013p8}. We therefore refer the reader to this paper for a detailed discussion of the observational procedures and reduction steps, such as optimal extraction of spectra, calculation of Stokes parameters, and continuum normalisation of the resulting spectra.

\section{Stellar parameters}\label{sec:fund}

An initial estimate of the effective temperature of HD\,119419 was obtained from the study by \citet{Kochukhov2006p763} who gave $T_\mathrm{eff} = 11150\pm400$\,K, while an estimate of the surface gravity $\log g =4.45 \pm 0.12$ was adopted from \citet{North1998p181}. Using these estimates, we computed a model atmosphere for $T_\mathrm{eff} = 11150$\,K, $\log g = 4.45$, and metallicity $[M/H] = +1.0$\,dex with the help of the \llm\ code \citep{Shulyak2004p993}. This model atmosphere code is based upon detailed line opacity calculations using abundances of individual chemical elements and incorporates a detailed treatment of Zeeman splitting and polarised radiative transfer \citep{Khan2006}. In these calculations we adopted a radial magnetic field with the strength of 18\,kG, which corresponds to the phase-averaged mean field modulus of HD\,119419 predicted by the field topology model of \citet{Bagnulo1999p158}. We used the updated version of the Vienna Line Database \citep[VALD3,][]{Kupka1999p119,Ryabchikova2015p54005} as a source of the atomic line data. 

Using the \llm\ model atmosphere, we determined abundances for about a dozen most important chemical elements, including He, Si, and several Fe-peak and rare-earth elements, by fitting the theoretical spectrum to the phase-averaged intensity spectrum (Stokes~$I$) of HD\,119419. Theoretical spectra were computed with the {\sc Synmast} code \citep{Kochukhov10p5}. The micro- and macroturbulent velocities were set to zero in all calculations because the strong magnetic field of HD\,119419 is expected to suppress all convective motions in the stellar atmosphere.

\begin{figure*}[!th]
  \centering
  \includegraphics[width=\textwidth]{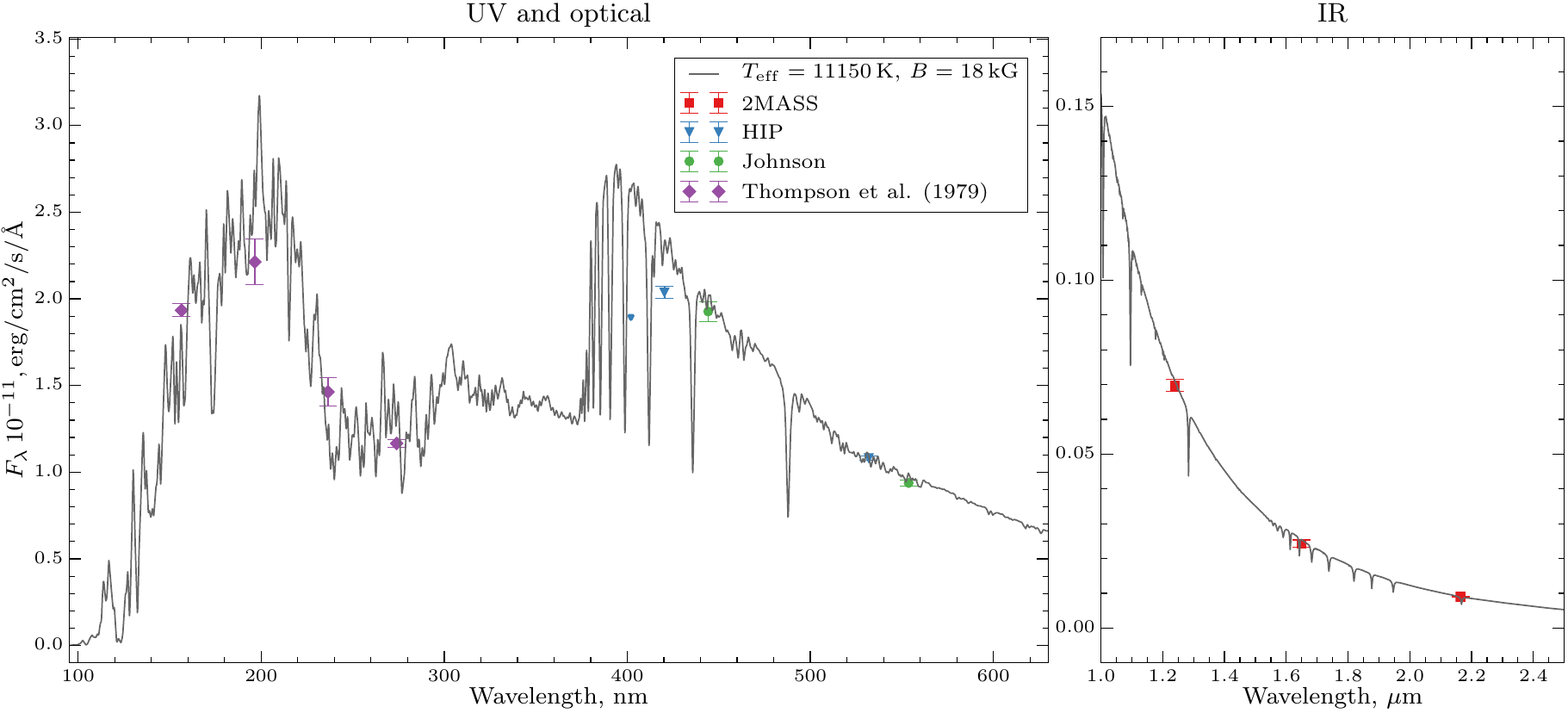}
  \caption{Comparison between the theoretical and observed spectrophotometry of HD\,119419. The light grey curve shows the theoretical SED for a model atmosphere with $T_\mathrm{eff}=11150$\,K, $\log g=4.30$, including the effects of individual non-solar abundances of HD\,119419 and a magnetic field with strength $\langle B \rangle = 18$\,kG. The photometric observations cover the UV (blue lines), optical (red circles) and near-IR (green squares) spectral regions.}\label{fig:sed}
\end{figure*}

In the next step we computed a grid of model atmospheres and spectral energy distributions (SEDs) with $T_\mathrm{eff} = 10000$--$12000$\,K and $\log g = 4.0$--4.5 in a 200\,K increment for the effective temperature and 0.1\,dex increment for the surface gravity. These calculations used individual abundances for the most important atmospheric opacity contributors and solar abundances for other elements. We then compared the calculated SEDs to the archival photometric measurements in the UV by \citet{Thompson1978}, Hipparcos and Johnson photometry measurements in the optical, and near-IR 2MASS observations (taken from the {\sc Simbad} data base). Largely due to the lack of detailed spectrophotometric observations in the optical and ultraviolet, such as the ones provided by \citet{Adelman1989p221} and the IUE satellite, it was not possible to find a better matching temperature than the one currently present in the literature. Nonetheless, we could confirm that the current $T_\mathrm{eff}=11150$~K provides a satisfactory description of the available observations as shown in Fig.~\ref{fig:sed}. Changing the effective temperature by about 300\,K produced flux distributions which were marginally worse than the one for the best-fitting effective temperature. 

During the flux fitting procedure we adopted zero interstellar reddening for HD\,119419, following \citet{Lipski2008p481} who suggested that stars closer than 100\,pc do not experience significant interstellar absorption. The Hipparcos parallax of HD\,119419 is $\pi = 8.40 \pm 0.46$\,mas \citep{Leeuwen07p653}, resulting in a distance of 119$\pm$7\,pc, which does not exceed 100\,pc by much.

\begin{figure}
  \centering
  \includegraphics[width=0.5\textwidth]{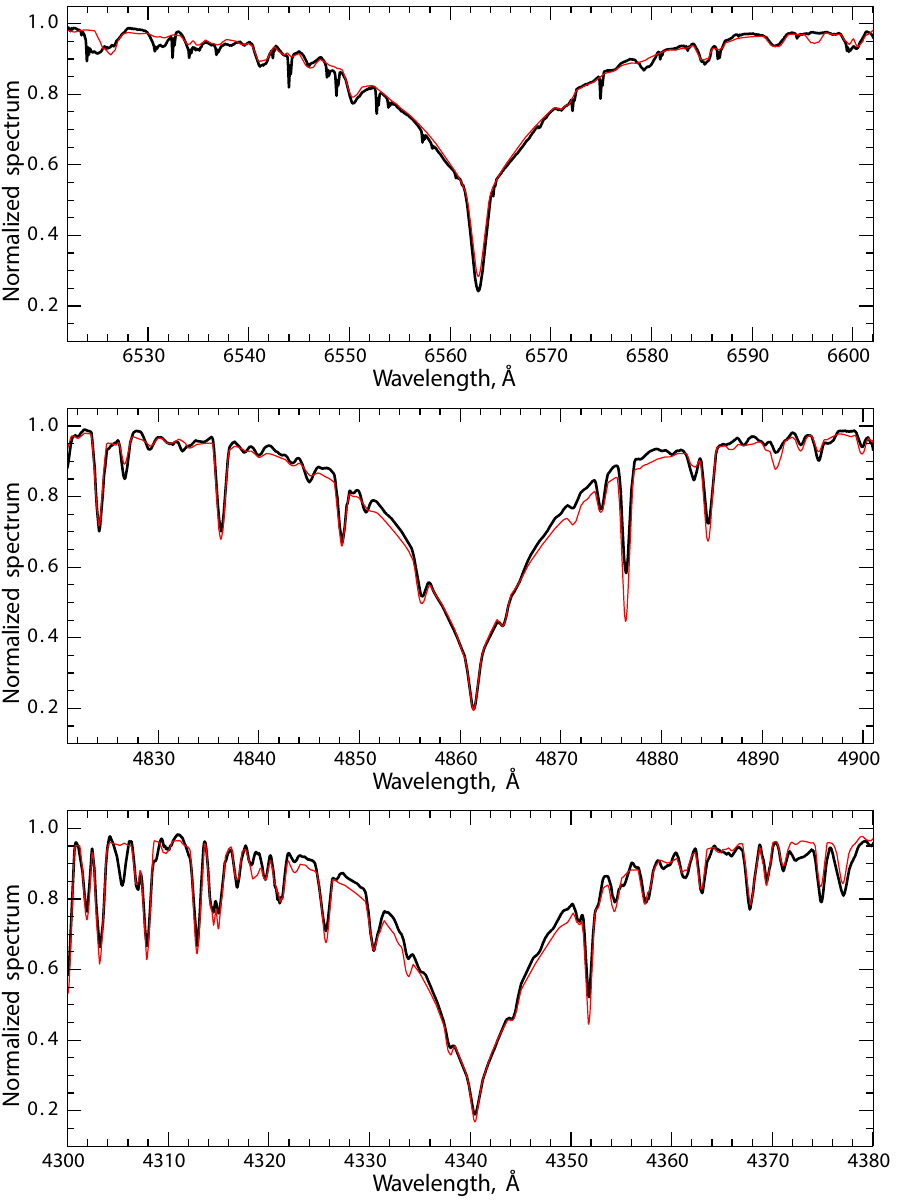}
  \caption{Comparison between the observed (thick black curve) and computed (thin red curve) profiles of the H$\alpha$, H$\beta$, and H$\gamma$ lines in the average spectrum of HD\,119419.}
  \label{fig:hlines}
\end{figure}

We constrained the surface gravity by fitting synthetic spectra to the observed profiles of the hydrogen Balmer lines. The $\log g$ values required to fit the H$\alpha$ line, on the one hand, and the H$\beta$ and H$\gamma$ lines, on the other hand, are slightly different. Specifically, H$\alpha$ favours a value of $\log g$ closer to 4.3, while H$\beta$ and H$\gamma$ are better fitted with a slightly lower value of 4.2. This discrepancy could be related to the difficulty of establishing an accurate continuum level in the blue part of the spectra. Therefore, we adopted $\log g = 4.3$ inferred from H$\alpha$ as the final value for the surface gravity of HD\,119419. The comparison between the observed and synthetic profiles of H$\alpha$, H$\beta$, and H$\gamma$ is presented in Fig.~\ref{fig:hlines}.

We used the parallax of the star, $\pi = 8.40 \pm 0.46$\,mas, together with the angular diameter yielded by the SED fitting procedure to derive the stellar radius, $R=1.8 \pm 0.2$\,$R_\odot$. This radius value agrees well with the determination by \citet[][1.9$\pm$0.3\,$R_\odot$]{North1998p181} who estimated the stellar radius on the basis of Geneva photometric calibration and Hipparcos parallax, and the value by \citet[][$2.1\pm0.1$\,$R_\odot$]{Bagnulo1999p158} inferred from fitting the phase curves of the longitudinal field, crossover and quadratic field measurements. We redetermined the luminosity of HD\,119419 using the new values of the effective temperature and the radius and found it to be $L=42 \pm 10$\,$L_\odot$, which is in good agreement with the determinations by \citet[][$49\pm10$\,$L_\odot$]{North1998p181} and \citet[][$44\pm8$\,$L_\odot$]{Kochukhov2006p763}.

Finally, we estimated the projected rotational velocity to be $v_{\rm e} \sin i = 27 \pm 3$\,\kms\ by fitting the Stokes~$I$ profiles of the magnetically insensitive Fe and Cr lines that were relatively free from blends by spectral lines of other chemical elements. Our $v_{\rm e} \sin i$ value agrees well with the determinations by \citet[][$27.7\pm5.2$\,\kms{}]{Bagnulo1999p158} and \citet[][$34.8\pm6.9$\,\kms{}]{Mathys1995p746}.

The stellar parameters of HD\,119419 determined in this paper are summarised in Table~\ref{tab:fundpars}.

\begin{table}
  \caption{Parameters of HD\,119149.}
  \begin{tabular}{ll}
  \hline
  \hline
  Parameter & Value \\
  \hline
$T_\mathrm{eff}$\,(K) & $11150 \pm 300$ \\
$\log g$ & $4.3\pm0.1$\\
$R/R_\odot$ & $1.8\pm0.2$\\
$L/L_\odot$ & $42\pm10$ \\
$v_{\rm e} \sin i$\, (\kms{}) & $27 \pm 3$ \\
  \hline
  \end{tabular}
  \label{tab:fundpars}
\end{table}

\subsection{Mean abundances}

Using the final model atmosphere parameters established above, we determined the abundances of elements for which we could find distinct  unblended lines that were suitable for abundance analysis. For other elements we assumed solar abundances, which were taken from \citet{Asplund2009p481}. A few elements lacked distinct lines, but yielded features much deeper than what is observed when we used solar abundances for the spectrum synthesis. For these elements we decreased the abundances until the synthetic and observed spectra matched as close as possible. We consider these estimates to be approximate upper limits. For example, the neutral He line at $\lambda$ 5875.6\,\AA\ is too strong with a helium abundance higher than $\log(N_{\rm He}/N_\mathrm{tot})$\,=\,$-$2.5\,dex.

Apart from silicon with its many \ion{Si}{ii} lines, most elements for which we could determine abundances belong to the Fe-peak and rare earth elements (REEs). The vast majority of iron-peak lines are those of singly ionised Fe, Cr, and Ti. Each of these elements has at least several lines with none or only a weak blending. We found that all of them have overabundances of 1--2\,dex relative to the solar values. 

Ce has the largest number of lines among REEs, but all of them are blended. Nevertheless, we were able to establish an upper abundance limit of roughly $\log(N_{\rm Ce}/N_\mathrm{tot})$\,=\,$-$7.0\,dex. The Ce lines in the theoretical spectra become too deep for a higher abundance of this element. Pr and Eu both have a couple of lines with weak blends, all showing abundances of about $-7.1$ dex. Er has one usable line at $\lambda$ 4735.554\,\AA, giving an upper limit of $-$8.5\,dex. La and Nd have the greatest number of lines after Ce, but unlike the latter, many of these lines have none or only weak blends. We found that both elements have abundances about 4\,dex higher than the solar values.

All of the abundances determined for HD\,119419 in this paper are reported in $\log (N_X / N_\mathrm{tot})$ units in Table~\ref{tab:abund}. In our abundance analysis we did not account for the hyperfine splitting or non-LTE effects, which can lead to systematic errors for some of the studied chemical elements.

\begin{table}
\caption{Mean atmospheric abundances of HD\,119419.}
\begin{tabular}{l l | l | l}
\hline \hline
Ion & $\log(N_\mathrm{ion}/N_\mathrm{tot})$ & $n$ & $\log(N/N_\mathrm{tot})_\odot$ \\
\hline
\ion{He}{i} & $< -2.5$ & & $-1.11$\\
\ion{Si}{ii} & $-3.1 \pm 0.2$ & 9 & $-4.53$\\
\ion{Ti}{ii} & $-5.1 \pm 0.2$ & 7 & $-7.09$\\
\ion{Cr}{ii} & $-3.9 \pm 0.2$ & 8 & $-6.40$\\
\ion{Fe}{ii} & $-3.3 \pm 0.3$ & 14 & $-4.54$\\
\ion{La}{ii} & $-6.5 \pm 0.1$ & 5 & $-10.94$\\
\ion{Ce}{ii} & $< -7$ & & $-10.49$\\
\ion{Pr}{iii} & $-7.1$ & 2 & $-11.32$\\
\ion{Nd}{iii} & $-6.7 \pm 0.1$ & 11 & $-10.62$\\
\ion{Eu}{ii} & $-7.12\pm 0.1$ & 2 & $-11.52$\\
\ion{Er}{iii} &  $< -8.5$ & & $-11.12$\\
\hline
\end{tabular}
\label{tab:abund}
\tablefoot{The first column identifies the ions for which we estimated the abundance given in the second column. The number of distinct lines analysed for each species is given in the third column. The solar abundances reported by \citet{Asplund2009p481} are given in the last column for comparison.}
\end{table}

\section{Mean polarisation profiles and integral magnetic observables}\label{sec:lsd}

\begin{figure*}
  \centering
  \includegraphics[width=0.49\textwidth]{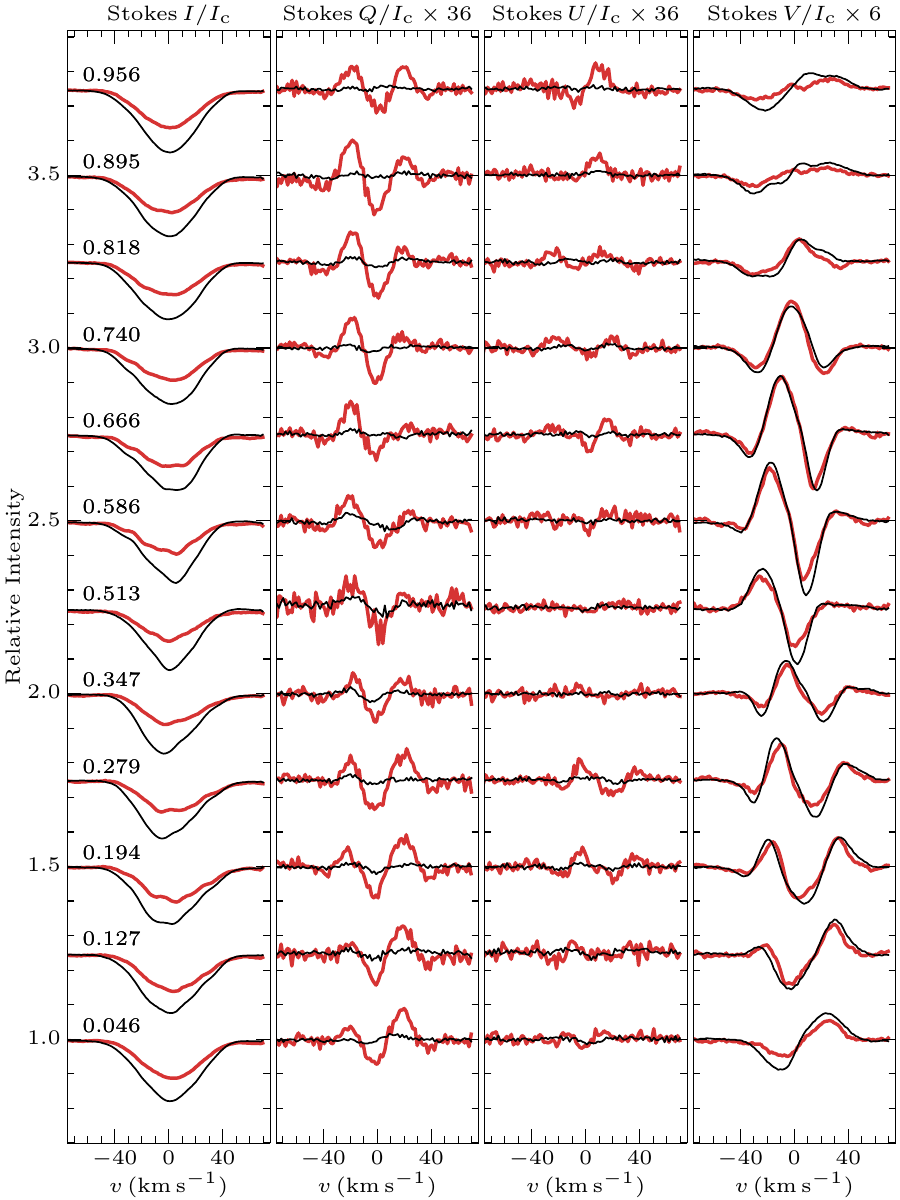}\hfill
  \includegraphics[width=0.49\textwidth]{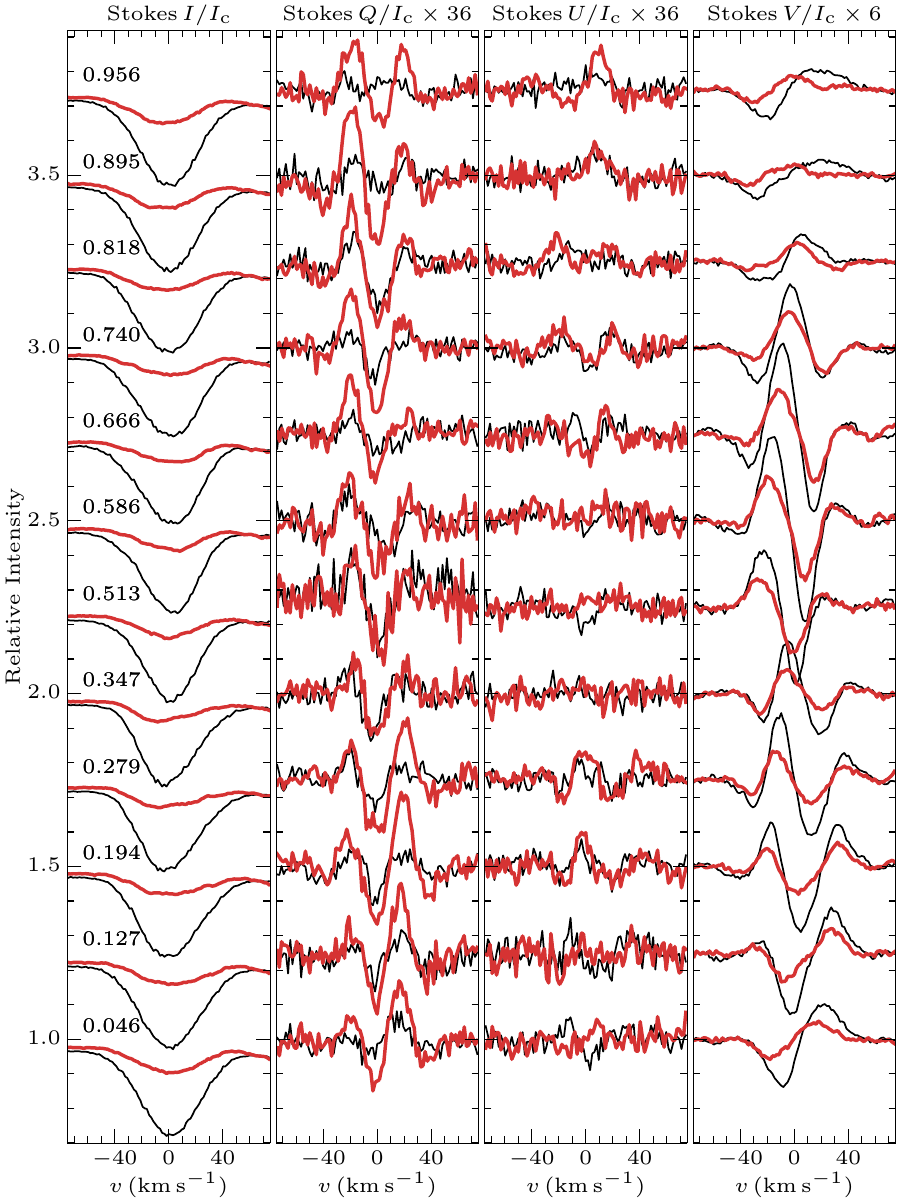}
  \caption{Mean Stokes~$IQUV$ line profiles of HD\,119419. The LSD Stokes profiles are shown in the left panel. The profiles of Fe-peak elements are shown with thin black lines, while the LSD profiles of REEs are plotted with thick red lines. The Stokes profiles computed by simple averaging of unblended lines are shown in the right panel. Profiles obtained by averaging \ion{Nd}{iii} lines are shown with thick red lines, while profiles obtained by averaging \ion{Fe}{ii}, \ion{Cr}{ii}, and \ion{Ti}{ii} lines are illustrated with thin black curves. The spectra are shifted vertically and are ordered according to the rotational phases indicated in the Stokes $I$ sub-panel of each plot.}
  \label{fig:mean-profiles}
\end{figure*}

It is possible to extract information about integral characteristics of a stellar magnetic field, such as the mean longitudinal field or the mean field modulus, directly from spectropolarimetric observations \citep[e.g.][]{Mathys2002}. The longitudinal magnetic field \bz\ can be inferred from both low- and high-resolution circular polarisation spectra. However, this is a non-trivial task for HD\,119419 because most of the observed line profiles are distorted by Zeeman splitting caused by a strong magnetic field and are blended by an amount that changes significantly with rotational phase. Additionally, the polarisation signal in individual lines is contaminated by random noise which further complicates the analysis. However, the magnetic field does influence a large number of spectral lines in roughly the same way. Having access to observations covering a wide wavelength range, it is possible to combine information from many lines and extract a clearer mean polarisation signal with a much higher signal-to-noise ratio. Least-squares deconvolution (LSD) is a technique developed to exploit this redundancy of the Zeeman information in individual lines \citep{Donati97p658,Kochukhov10p5}. This method assumes that a stellar spectrum can be represented as a linear superposition of mean profiles, scaled by a factor that depends on the line strength, magnetic sensitivity, and the central wavelength. For this study we employed the multi-profile version of the LSD method described by \citet{Kochukhov10p5} to the Stokes spectra of HD\,119419. This implementation of the LSD technique allows one to simultaneously extract several mean profiles corresponding to different groups of spectral lines.

The VALD3 database \citep{Ryabchikova2015p54005} was used as a source of the atomic line data necessary for compilation of an LSD line mask. We constructed the line masks by first removing all spectral lines affected by the hydrogen Balmer lines and contaminated by the telluric absorption, and lines with the intrinsic central depth less than 0.1 of the continuum intensity. From this line mask we also removed the lines of light elements as well as of all rare-earth elements with the exception of \ion{Nd}{iii}, \ion{La}{ii}, and \ion{Pr}{iii}, which have reliably determined abundances and many distinct lines in the spectrum of HD\,119419. The final line mask was comprised of 1883 lines belonging to Fe-peak elements and 206 REE lines (108 \ion{La}{ii}, 52 \ion{Pr}{iii}, and 46 \ion{Nd}{iii}). Using this mask we simultaneously extracted two sets of LSD profiles. These profiles were normalised using the following parameters: $\lambda_0 = 5000$\,\AA, depth $d_0 = 1$, and effective Land\'e factor $\bar{g}_0 = 1$. The resulting LSD Stokes profiles are shown in Fig.~\ref{fig:mean-profiles} (left panel).

Some surprising features are immediately obvious from this Figure. Although, the Stokes~$I$ LSD profiles of REEs are systematically weaker than the Stokes~$I$ profiles of Fe-peak elements, the Stokes~$QU$ LSD profiles of REEs show the opposite behaviour, while the Stokes~$V$ profiles of both groups of elements show similar amplitudes. This is an unusual behaviour because normally one expects the amplitude of the Stokes~$QU$ profiles of REEs to be proportional to the intensity. 

Given the unexpected behaviour of the amplitudes of the LSD Stokes profiles, we decided to investigate this result further. To this end, we compiled a set of \ion{Nd}{iii}, \ion{Fe}{ii}, \ion{Cr}{ii}, and \ion{Ti}{ii} lines with visible polarisation signatures and minimal blending in the observed spectra. We then used this line list to directly compute the average profiles of \ion{Nd}{iii} and Fe-peak lines without resorting to the LSD technique. The resulting mean Stokes profiles are shown in Fig.~\ref{fig:mean-profiles} (right panel). Unlike in the LSD method, none of the lines were weighted according to their depth, wavelength or magnetic sensitivity. Most of the Fe-peak lines are two to three times deeper than the Nd lines, so the relative intensity of the Nd profiles is even weaker than for the REE LSD profiles. Still, we can see that the amplitude of the \ion{Nd}{iii} Stokes $QU$ profiles matches or even surpasses the corresponding Fe-peak profiles. This result confirms that the REE (or at least \ion{Nd}{iii}) lines in the spectrum of HD\,119419 have unusually strong polarisation signatures compared to the Fe-peak elements. This might indicate unusually high-contrast horizontal abundance inhomogeneities or an unusual magnetic field geometry or both.

The LSD Stokes profiles obtained above can be used to derive quantitative characteristics of the magnetic field of HD\,119419, such as the longitudinal magnetic field \bz\ \citep{Mathys1991p121} and the net linear polarisation $P_Q$ and $P_U$ \citep{Wade00p851}. The former measures the strength of the disk-integrated magnetic field component parallel to the line of sight and can be calculated from the first order moment of the Stokes~$V$ profiles with the following formula:
\begin{equation}
\langle B_\mathrm{z} \rangle = -7.145 \cdot 10^6 \frac{\int v V(v)\,dv}{\lambda_0 \bar{g}_0 \int [1-I(v)]\,dv},
\end{equation}
where \bz\ is measured in Gauss and $\lambda_0$ in \AA. The quantities $\lambda_0$ and $\bar{g}_0$ are the constants used for the normalisation of the LSD profiles, that is, $\lambda_0 = 5000$\,\AA\ and $\bar{g}_0 = 1$.

The \bz\ measurements obtained from the LSD profiles of Fe-peak elements and REEs are presented in Table~\ref{tab:bz}. The corresponding uncertainties were obtained from the LSD profile error bars using the standard error propagation technique. Using the same technique, we calculated \bz\ from the null LSD profiles, which were derived by applying the same line masks and the multi-profile LSD method to the diagnostic null Stokes spectra. The resulting \bz\ values can be used as an indicator of contribution of spurious polarisation to longitudinal field measurements. This analysis showed that any spurious contribution to \bz\ is below the uncertainties of the measurements due to the finite S/N of the observations. 

We also attempted to measure the net linear polarisation from the Stokes~$Q$ and $U$ LSD profiles. This magnetic field characteristic can be calculated by integrating the signal in the respective profiles and normalising it to the equivalent width of the Stokes~$I$ profiles; see \citet{Wade00p851} and \cite{Kochukhov10p5} for more information about the interpretation of net linear polarisation measurements from LSD profiles. Our attempt to measure these quantities was unsuccessful. Despite clear linear polarisation signatures evident in individual lines and in mean profiles in Fig.~\ref{fig:mean-profiles}, the integral measurements were not significantly different from either zero or analogous measurements of the null profiles and did not exhibit a coherent phase dependence. This negative result is unsurprising given that the net linear polarisation is thought \citep{Wade00p851} to be linked, similar to the broad-band linear polarisation measured with photopolarimetry \citep{Landolfi1993p285}, to the differential saturation of $\pi$ and $\sigma$ components. This mechanism requires strong, saturated spectral lines and therefore is effective only in relatively cool Ap stars.

\begin{table}[!t]
  \caption{Longitudinal magnetic field measurements obtained from the LSD Stokes~$V$ profiles of HD\,119419.}
  \centering
  \begin{tabular}{ccrr}
  \hline
  \hline
& & \multicolumn{2}{c}{$\langle B_\mathrm{z} \rangle$ (G)}\\
HJD\,(2\,455\,000$+$) & Phase    & Fe-peak & REE \\
\hline
1350.6960 & 0.026 &  $-$1954 $\pm$ 15&  $-$1903 $\pm$ 61\\
1345.7042 & 0.107 &  $-$1574 $\pm$ 19&  $-$1572 $\pm$ 80\\
1015.6141 & 0.178 &  $-$1068 $\pm$ 21&  $-$1226 $\pm$ 95\\
1348.7021 & 0.259 &  $-$393 $\pm$ 23&  $-$607 $\pm$ 106 \\
1018.6095 & 0.330 &  116 $\pm$ 19&  27 $\pm$ 93 \\
1346.7097 & 0.493 &  1266 $\pm$ 22&  1767 $\pm$ 95 \\
1016.6307 & 0.569 &  1374 $\pm$ 28&  2347 $\pm$ 131\\
1349.7079 & 0.646 &  585 $\pm$ 29&  1429 $\pm$ 128\\
1019.6244 & 0.720 &  $-$515 $\pm$ 19&  43 $\pm$ 86 \\
1014.6345 & 0.801 &  $-$1397 $\pm$ 12&  $-$1078 $\pm$ 51\\
1347.7024 & 0.875 &  $-$1757 $\pm$ 12&  $-$1544 $\pm$ 49\\
1017.5944 & 0.939 &  $-$1895 $\pm$ 12&  $-$1784 $\pm$ 52\\
\hline
  \end{tabular}
  \label{tab:bz}
  \tablefoot{The first and second columns list the heliocentric JD and corresponding rotational phase of the Stokes $V$ observations. The next two columns give \bz\ measurements obtained from the REE and Fe-peak LSD Stokes~$V$ profiles.}
\end{table}

\subsection{Rotational period}\label{sec:period}
We employed longitudinal magnetic field measurements to revise the value of the rotation period of HD\,119419. This set of \bz\ measurements was produced by combining our HARPSpol Fe-peak \bz\ determinations with the \ion{Fe}{ii} line measurements from the literature \citep{Mathys1994p547,Mathys97p475}. These data were complemented by the photopolarimetric longitudinal field measurements from the wings of H$\beta$ lines reported by \citet{Bohlender1993p355} and \citet{Thompson1987p219}. 

The entire data set is comprised of 42 measurements in total and spans over a time period of close to 30 years. We used the period $P_\mathrm{rot} = 2.6009$\,days reported by \citet{Mathys97p475} as a starting guess, and adopted the zero point from the paper by \citet{Mathys1991p121}. We found that using four harmonic components results in a good overall fit to the \bz\ data. With the zero point from \citet{Mathys1991p121} and the revised rotational period, the final ephemeris is given by
\begin{equation}
\mathrm{HJD} ( {\langle B_\mathrm{z} \rangle}_\mathrm{min} ) = 2\,446\,512.596+2.60059(1)\cdot \mathrm{E}.
\end{equation}

\begin{figure}
\centering
\includegraphics[width=0.49\textwidth]{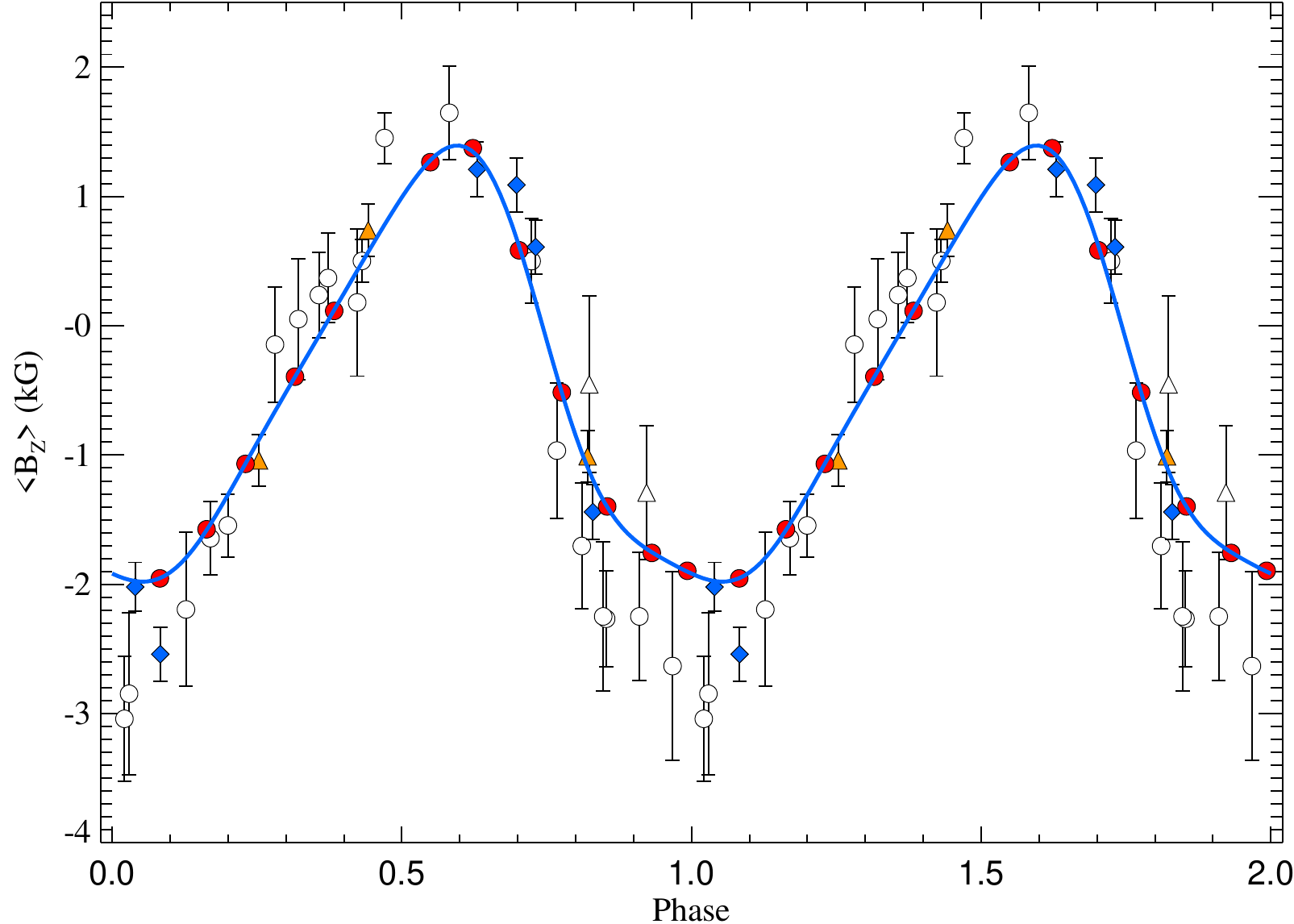} \label{fig:period}
\caption{Variability of the mean longitudinal magnetic field of HD\,119419 with the rotational phase. White circles are from \citet{Mathys1994p547}, white triangles from \citet{Mathys97p475}, orange triangles from \citet{Thompson1987p219}, and blue diamonds from \citet{Bohlender1993p355}. Our data points are shown with red circles. The thick solid curve is the fourth-order Fourier fit to all longitudinal field measurements.}
\end{figure}

The variation of the mean longitudinal field of HD\,119419 phased with the improved rotational period is illustrated in Fig.~\ref{fig:period}. All phases in this paper are computed according to the ephemeris given above.

\section{Magnetic Doppler imaging}\label{sec:mdi}

\subsection{Surface mapping methodology}

We reconstructed the magnetic field topology and distributions of several chemical elements on the surface of HD\,119419 with the help of the Magnetic Doppler imaging code \invers{10} \citep{Kochukhov2002p868,Piskunov2002p736}. This code iteratively adjusts the surface magnetic field and the abundance maps of the studied chemical elements by comparing theoretical intensity and polarisation Stokes spectra to the observed ones for each rotational phase until a desirable fit is achieved. This paper follows the same methodology of the magnetic inversions as was repeatedly applied by previous similar MDI studies \citep[e.g.][]{Silvester2014p182,Rusomarov2015p123,Rusomarov2016p138}.

The version of the \invers{10} code that we applied in this study uses a spherical harmonic description of the surface magnetic field vector distribution introduced by \citet{Donati06p629} and revised by \citet{Kochukhov2014p83}. In this framework, the magnetic field is parameterised by a set of the harmonic coefficients $\alpha_{\ell m}$, $\beta_{\ell m}$, and $\gamma_{\ell m}$, which correspond to the poloidal radial, poloidal horizontal, and toroidal horizontal field components, respectively. In this study we truncated the harmonic expansion at $\ell_\mathrm{max} = 10$ since we found that including higher order harmonics did not improve the fit to the observations.

Following \citet{Kochukhov2014p83}, we limit the contribution of the higher-order spherical harmonic modes by using a penalty function $\sum_{\ell, m} \ell^2 (\alpha_{\ell m} + \beta_{\ell m} + \gamma_{\ell m})^2$. For the reconstruction of the chemical abundance maps, we applied the standard Tikhonov regularisation \citep{Piskunov2002p736,Kochukhov2017}, which yields the smoothest possible abundance maps as permitted by the observations.

\subsection{Spectral line selection}\label{ssec:mdi:lines}

For the Magnetic Doppler imaging of HD\,119419 we selected a set of 7~\ion{Fe}{ii}, 3~\ion{Cr}{ii}, and 4~\ion{Nd}{iii} lines. These lines were chosen based on the presence of clear signatures in Stokes~$V$ and minimal blending from spectral lines of other chemical elements. We attempted to include both lines with moderate to weak magnetic sensitivity (effective Land\'e factor $\bar{g} \le 1$) and lines with strong sensitivity to the magnetic field ($\bar{g} > 1$). The latter are excellent for the magnetic field mapping owing to their strong Zeeman signatures, while the former are useful for performing an accurate abundance mapping and constraining the projected rotational velocity $v_{\rm e} \sin i$. The list of primary spectral features used for MDI is presented in Table.~\ref{tab:linelist}. For each line relevant minor blends were included in the modelling of corresponding wavelength regions. The atomic data for all the lines were extracted from the VALD3 data base.

\begin{table}
\caption{Spectral lines used in magnetic Doppler Imaging of HD\,119419.}\label{tab:linelist}
\begin{tabular}{lcccc}
\hline
\hline
Ion & $\lambda$\,(\AA)& $E_\mathrm{i}$\,(eV)& $\log gf$& $\bar{g}$\\
\hline
\ion{Fe}{ii} & 4491.397& 2.856& $-2.700$& 0.42\\
\ion{Fe}{ii} & 4508.280& 2.856& $-2.250$& 0.51\\
\ion{Cr}{ii} & 4588.199& 4.071& $-0.627$& 1.06\\
\ion{Cr}{ii} & 4592.049& 4.074& $-1.221$& 1.21\\
\ion{Fe}{ii} & 4731.448& 2.891& $-3.000$& 0.66\\
\ion{Nd}{iii} & 4759.526& 0.631& $-1.150$& 1.46\\
\ion{Nd}{iii} & 4769.622& 0.296& $-1.800$& 0.87\\
\ion{Cr}{ii} & 4812.337& 4.071& $-1.960$& 1.50\\
\ion{Fe}{ii} & 4923.921& 2.891& $-1.320$& 1.70\\
\ion{Nd}{iii} & 4927.488& 0.461& $-0.800$& 1.18\\
\ion{Nd}{iii} & 4942.673& 0.631& $-1.230$& 1.78\\
\ion{Fe}{ii} & 5018.436& 2.891& $-1.220$& 1.94\\
\ion{Nd}{iii} & 5050.695& 0.296& $-1.060$& 1.17\\
\ion{Fe}{ii} & 5169.028& 2.891& $-1.125$& 1.33\\
\hline
\end{tabular}
\tablefoot{The columns give the ion, central wavelength $\lambda$, excitation potential of the lower atomic level $E_i$, the oscillator strength $\log gf$, and the effective Land\'e factor $\bar{g}$.}
\end{table}

It is worth mentioning that in contrast to other stars for which we have performed MDI in the past, for example, HD\,125248 \citep{Rusomarov2016p138} and HD\,24712 \citep{Rusomarov2015p123}, HD\,119419 presents only a handful of spectral lines that show clear polarisation signatures in the Stokes $QU$ parameters. As is the case for many other Ap stars \citep{Wade00p823}, the strong \ion{Fe}{ii}~4923\,\AA\ and 5018\,\AA\ lines exhibit the most prominent linear polarisation signatures in the spectrum of HD\,119419. In addition, the Stokes $I$ profiles of these lines show multiple narrow components, most likely manifesting partially resolved Zeeman splitting. However, these lines are saturated over a large fraction of the stellar surface which complicates their use for abundance mapping. To limit their effect on the abundance mapping and $v_{\rm e} \sin i$ determination, we included intermediate-strength \ion{Fe}{ii} lines 4491\,\AA, 4508\,\AA, and 4731\,\AA,\ even though they did not show distinct Stokes~$QU$ signatures. In the case of Cr, we included the \ion{Cr}{ii} line 4812\,\AA, which shows an interesting modulation with the rotational phase and fine structure in the Stokes~$I$ profiles, and the \ion{Cr}{ii}~4592\,\AA\ line, which was the only other unblended Cr absorption feature we could find in the spectrum.

\subsection{Determination of $v_{\rm e} \sin i$ and orientation of the rotational axis}

The MDI technique requires the knowledge of the orientation of the stellar rotational axis relative to the observer and the projected rotational velocity $v_{\rm e} \sin i$. The three-dimensional orientation of the rotational axis is defined by the inclination angle $i$, which is a tilt of the rotational axis relative to the line of sight of the observer, and the azimuthal angle $\Theta$, which determines the orientation of the rotational axis in the plane of the sky. The inclination angle $i$ can have values between $0^\circ$ and $180^\circ$. The azimuth angle $\Theta$ has values in the $[0^\circ, 360^\circ]$ range. However, because the Stokes~$QU$ parameters depend on the trigonometric functions of $2\Theta$, there is an inherent ambiguity between $\Theta$ and $\Theta+180^\circ$.

We employed the \invers{10} code to investigate the nuisance parameter space in order to find a set of angles and $v_{\rm e} \sin i$ yielding the best fit to observations. \citet{Kochukhov2002p868} showed that an incorrect value of $v_{\rm e} \sin i$ leads to an increase of the $\chi^2$ and appearance of the characteristic axisymmetric artefacts in the reconstructed abundance maps. Thus, it is possible to refine the value of $v_{\rm e} \sin i$ by optimising the fit to the Stokes~$I$ profiles. Likewise, adopting an incorrect inclination or azimuth angle leads to an increase of the $\chi^2$ of the fit to the observed Stoke~$Q$ and $U$ profiles.

We calculated the initial value of the inclination angle, $i=50^\circ \pm 10^\circ$, by employing the oblique rotator relation taking into account the stellar radius $R=1.8\pm0.2\,R_\odot$, the projected rotational velocity $v_{\rm e} \sin i = 27\pm3$\,\kms\ (Sect.~\ref{sec:fund}), and the rotational period $P_\mathrm{rot} = 2.60059$\,d (Sect.~\ref{sec:period}). We then calculated a series of MDI inversions changing $i$ between $30^\circ$ and $70^\circ$ and $\Theta$ between $0^\circ$ and $180^\circ$ with $10^\circ$ steps in both angles. In total, we performed 90 MDI inversions for all the combinations of $i$ and $\Theta$. We found that the minimum $\chi^2$ corresponds to $i=60^\circ$ and $\Theta=170^\circ$. 

Then, we optimised the projected rotational velocity by performing MDI inversions for values of $v_{\rm e} \sin i$ between 20 and 31\,\kms\ and examining the $\chi^2$ of the fit to the observed Stokes~$I$ profiles for the REE and Fe-peak lines separately. Both groups of lines indicated a $\chi^2$ minimum around 24--25~\kms. We therefore adopted $v_{\rm e} \sin i = 25\pm1$\,\kms\ for the subsequent analysis.

The revised inclination and $v_{\rm e} \sin i$ correspond to the stellar radius $R=1.5\pm0.2$\,$R_\odot$, which is smaller than the value found in Sect.~\ref{sec:fund}. Nevertheless, both determinations are compatible within error bars. As an additional cross-check, we also compared the Stokes~$Q$ and $U$ profile fit for $i=50$\degr\ and $i=60$\degr, finding the former to be inferior. Based on these arguments, we adopted $v_{\rm e} \sin i = 25\pm1$\,\kms, $i=60^\circ \pm 10$\degr, and  $\Theta=170\pm10$\degr\ for the rest of this study. 

\subsection{Magnetic field topology}

\begin{figure*}
  \centering
  \includegraphics[width=\textwidth]{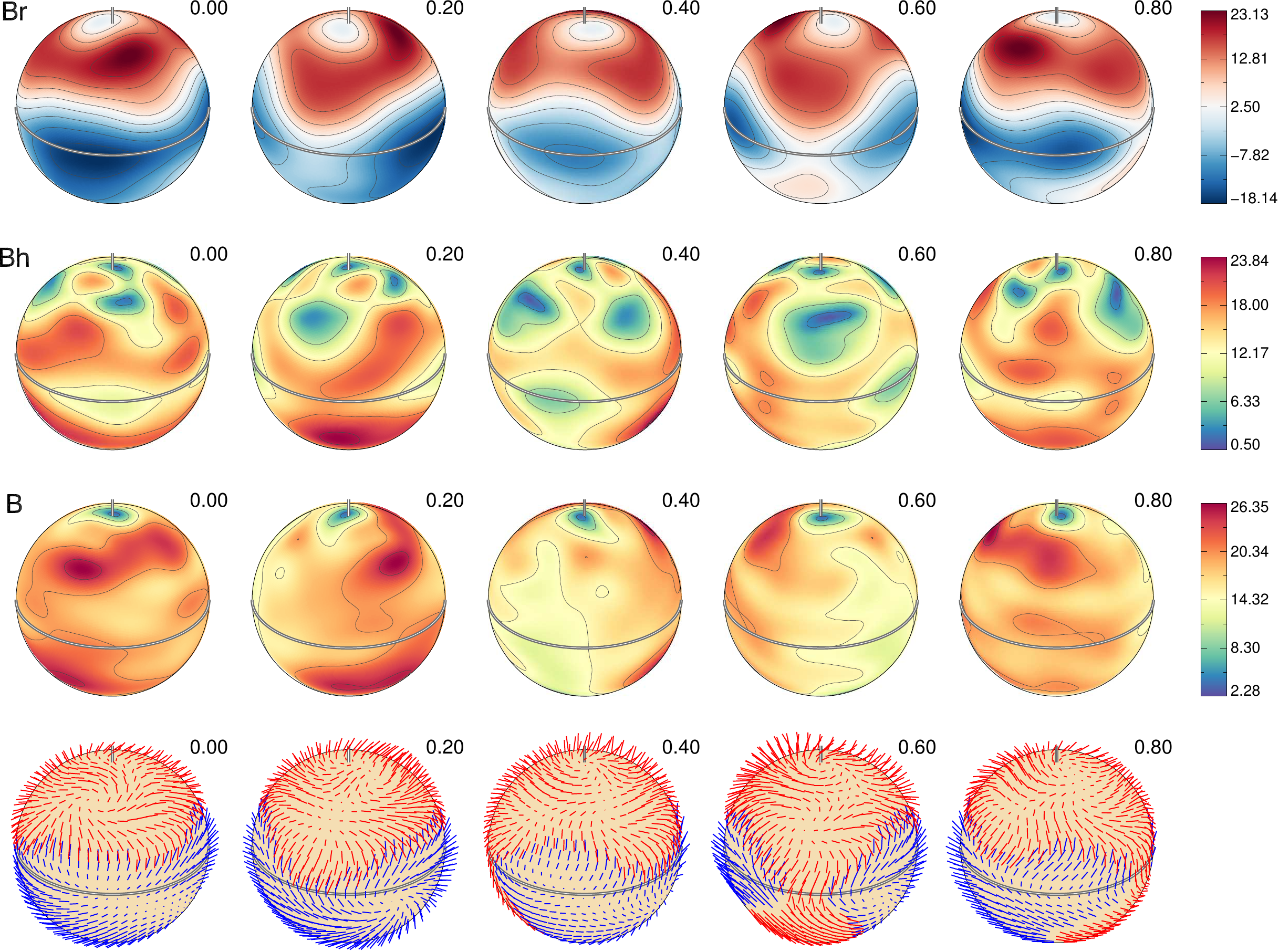}
  \caption{Distribution of the magnetic field on the surface of HD\,119419 derived from simultaneous MDI analysis of Fe, Cr, and Nd lines. The plots show the distribution of the radial magnetic field (top row), horizontal field (second row), field modulus (third row), and field orientation (fourth row). The colour bars on the right indicate the field strength in kG. The contours are plotted with 5\,kG increments. The arrow length in the bottom plot is proportional to the field strength. The star is shown at five rotational phases, indicated next to each spherical plot. The thick line and the vertical bar indicate positions of the rotational equator and the pole respectively.}\label{fig:mdi:mf}
\end{figure*}

\begin{figure*}
  \centering
  \includegraphics[width=\textwidth]{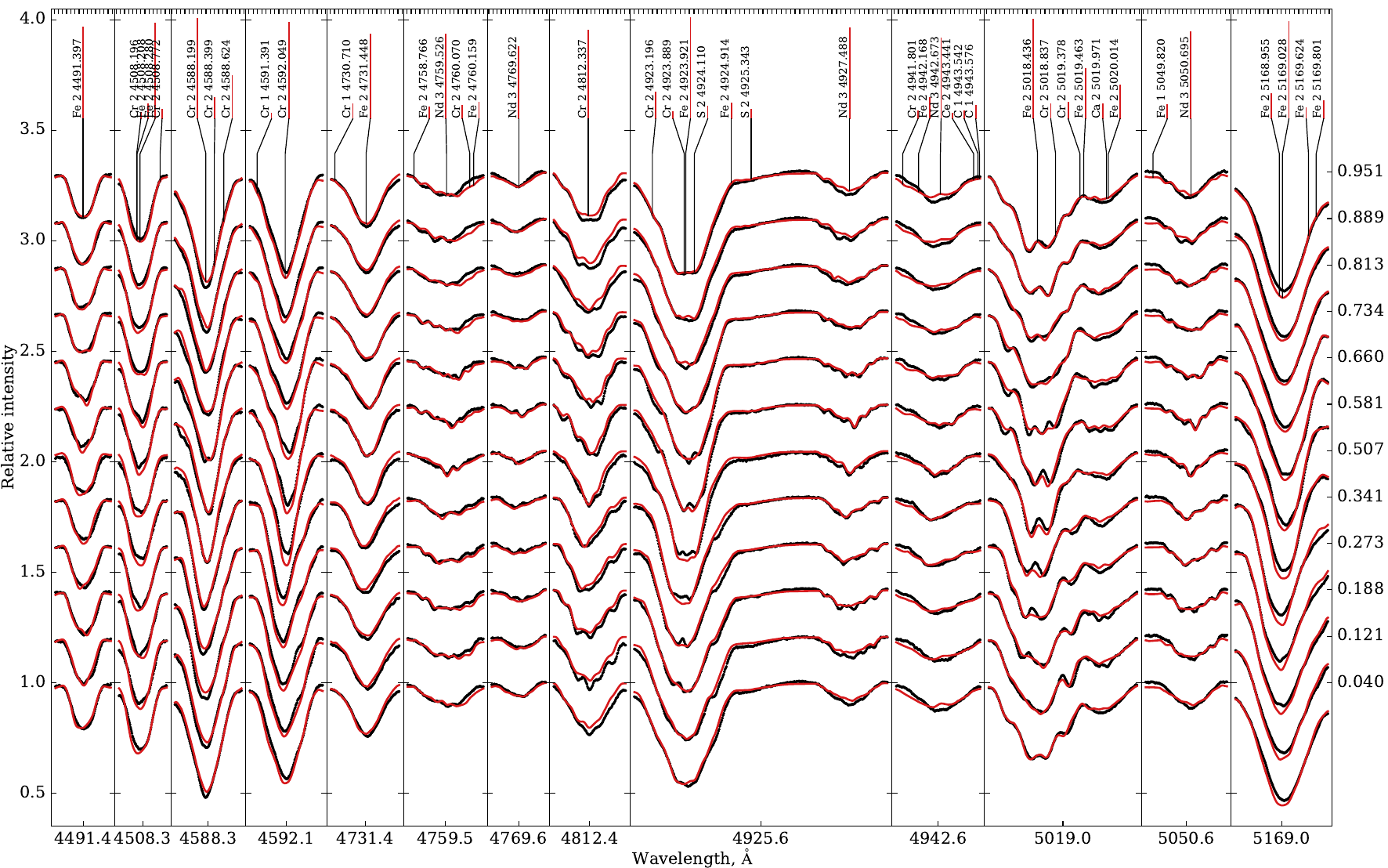}
  \includegraphics[width=\textwidth]{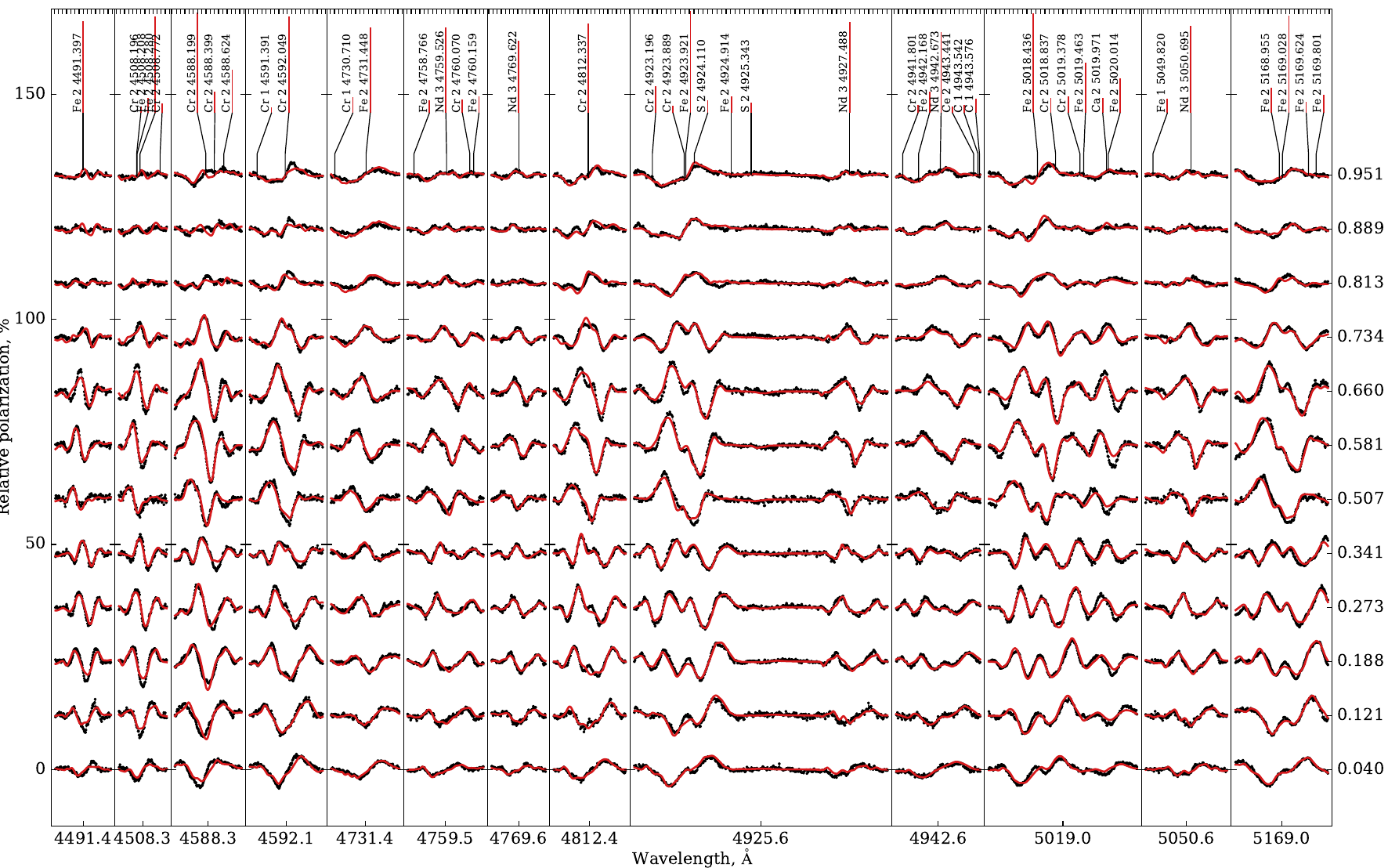}
  \caption{Comparison of the observed (symbols connected with black lines) and synthetic (thin red lines) Stokes~$I$ (upper panel) and $V$ (lower panel) profiles calculated for the final magnetic field and abundance maps. The distance between two horizontal tick marks on the upper axis of each panel is 0.1\,\AA, indicating the wavelength scale. Rotational phases are given on the right of both panels. The spectra are shifted vertically and order according to the rotational phase.}\label{fig:mdi:StokesIV}
\end{figure*}

\begin{figure*}
  \centering
  \includegraphics[width=\textwidth]{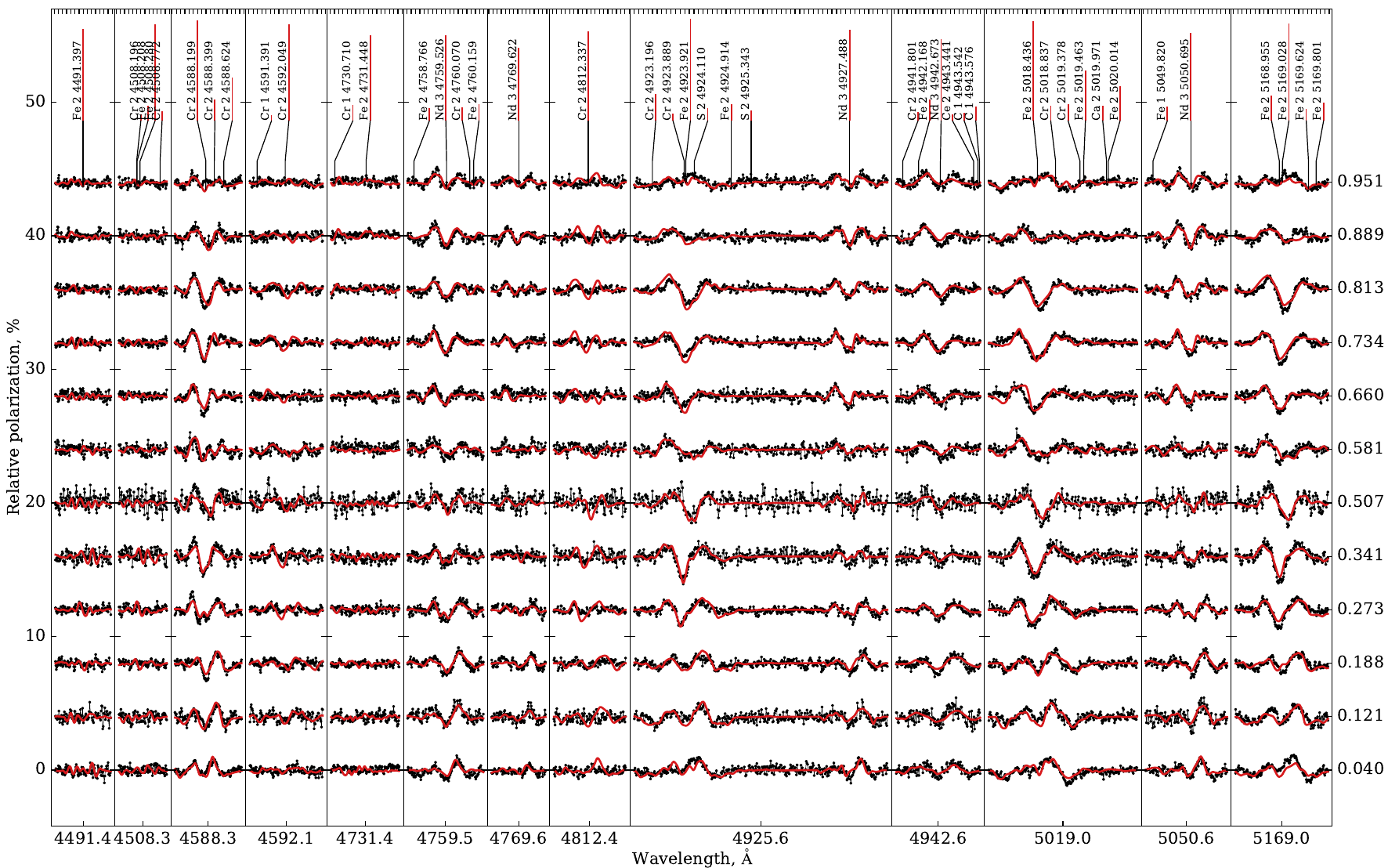}
  \includegraphics[width=\textwidth]{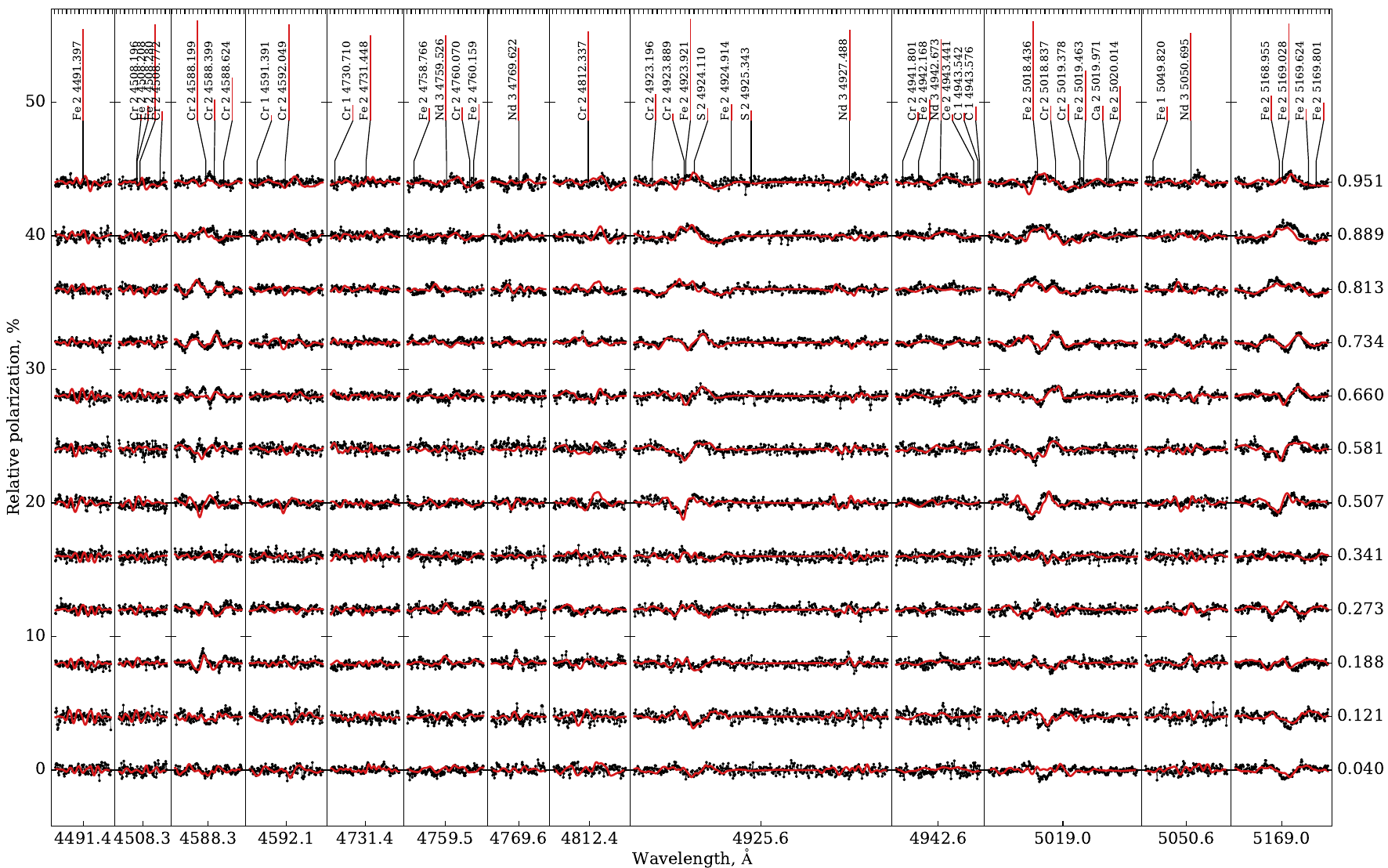}
  \caption{Same as Fig.~\ref{fig:mdi:StokesIV} for the Stokes~$Q$ and $U$ profiles.}\label{fig:mdi:StokesQU}
\end{figure*}

We derived the magnetic field topology of HD\,119419 by fitting the phase variations of the Stokes~$IQUV$ profiles of the Fe, Cr, and Nd lines listed in Table~\ref{tab:linelist}. The inferred surface magnetic field distribution is illustrated in Fig.~\ref{fig:mdi:mf}. In this Figure, different rows present spherical projections of the radial field component, magnetic field modulus, the strength of the horizontal field, and orientation of the magnetic field vectors. Figures~\ref{fig:mdi:StokesIV} and~\ref{fig:mdi:StokesQU} compare the observed and synthetic spectra for the Stokes~$IV$ and $QU$ parameters, respectively. These Figures indicate that, with a few exceptions, the inversions code achieves a very good description of both circular and linear polarisation profiles, and reproduces partially resolved Zeeman splitting seen in the cores of several \ion{Fe}{ii} and \ion{Nd}{iii} lines.

The results of the magnetic inversions indicate that the field structure of HD\,119419 is quasi-dipolar only on the largest spatial scales. As indicated by Fig.~\ref{fig:mdi:mf} (top and bottom rows), there are two distinct areas where the field is predominantly of negative polarity (inward oriented field vectors) or positive polarity (outward oriented field vectors). However, there are obvious deviations from a pure axisymmetric dipolar geometry. This can be most easily seen around phase 0.6, when the positive field polarity region extends towards the Southern rotational pole. We also find some asymmetry in the field strength between the areas of the maximum positive and negative magnetic field.

Using the MDI magnetic field map of HD\,119419, we estimated the disk-integrated mean field modulus and found that it varies between 19.7 and 15.3~kG. The phase-averaged mean field modulus is $\langle B \rangle = 17.8$\,kG. This value matches fairly well the figure of $\langle B \rangle = 18$\,kG predicted by the field geometry model of \citet{Bagnulo1999p158}, but is somewhat higher than $\langle B \rangle = 14.7$\,kG implied by the updated dipole-quadrupole model published by \citet{Bagnulo2002p1023}. However, other details of either of these dipole plus quadrupole topologies, such as the overall field geometry and the local field strength reaching 40--53~kG, are incompatible with the MDI results. Similar disagreements can be noted also for the off-centred dipolar model proposed by \citet{Glagolevskii2001}.

\begin{figure}
  \centering
  \includegraphics{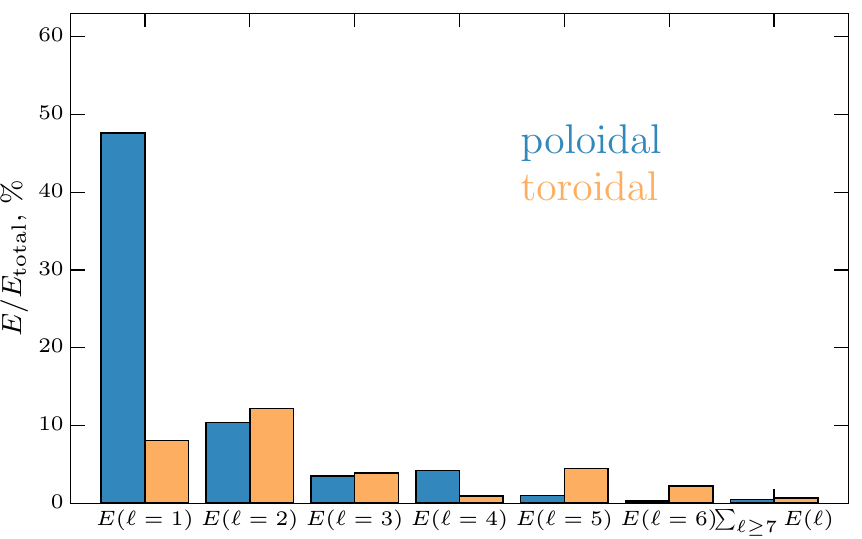}
  \caption{Relative energy of the poloidal and toroidal harmonic components of the surface magnetic field of HD\,119419. The energy of the poloidal and toroidal components is shown in dark (blue) and light (orange), respectively. The last two bars represent the sum of energies for $\ell \geq 7$ modes.}\label{fig:mdi:energy}
\end{figure}

We analysed the contribution of different harmonic components to the total energy of the magnetic field of HD\,119419 and found that the field is mostly poloidal, with 68\% of the total magnetic energy. Nevertheless, the remaining 32\% of the magnetic energy concentrated in the toroidal field is significant and exceeds toroidal field strength found with MDI for some other Ap stars. The poloidal dipolar component dominates over all other components, contributing 48\% to the total field energy. The quadrupolar $(\ell=2)$ poloidal and toroidal field components contribute 10 and 12\% to the total field energy, respectively. It is worth mentioning that in the case of HD\,119419 the toroidal quadrupolar field component is slightly stronger than the corresponding dipolar component. The field harmonics with $\ell \ge 3$ contribute slightly less than 22\% to the total field energy. In Fig.~\ref{fig:mdi:energy} we illustrate the energies of the poloidal and toroidal components of the magnetic field as a function of the spherical harmonic angular degree $\ell$.

\begin{figure*}
        \centering
        \includegraphics{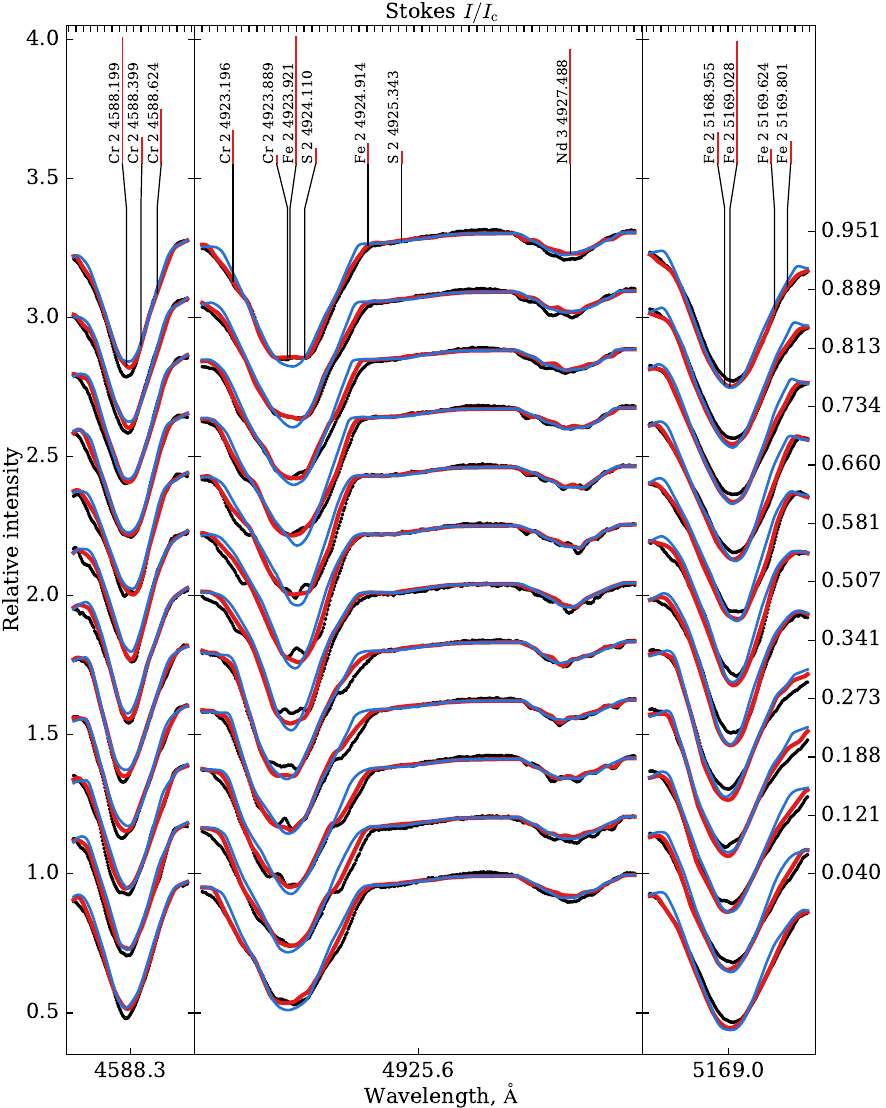}
        \includegraphics{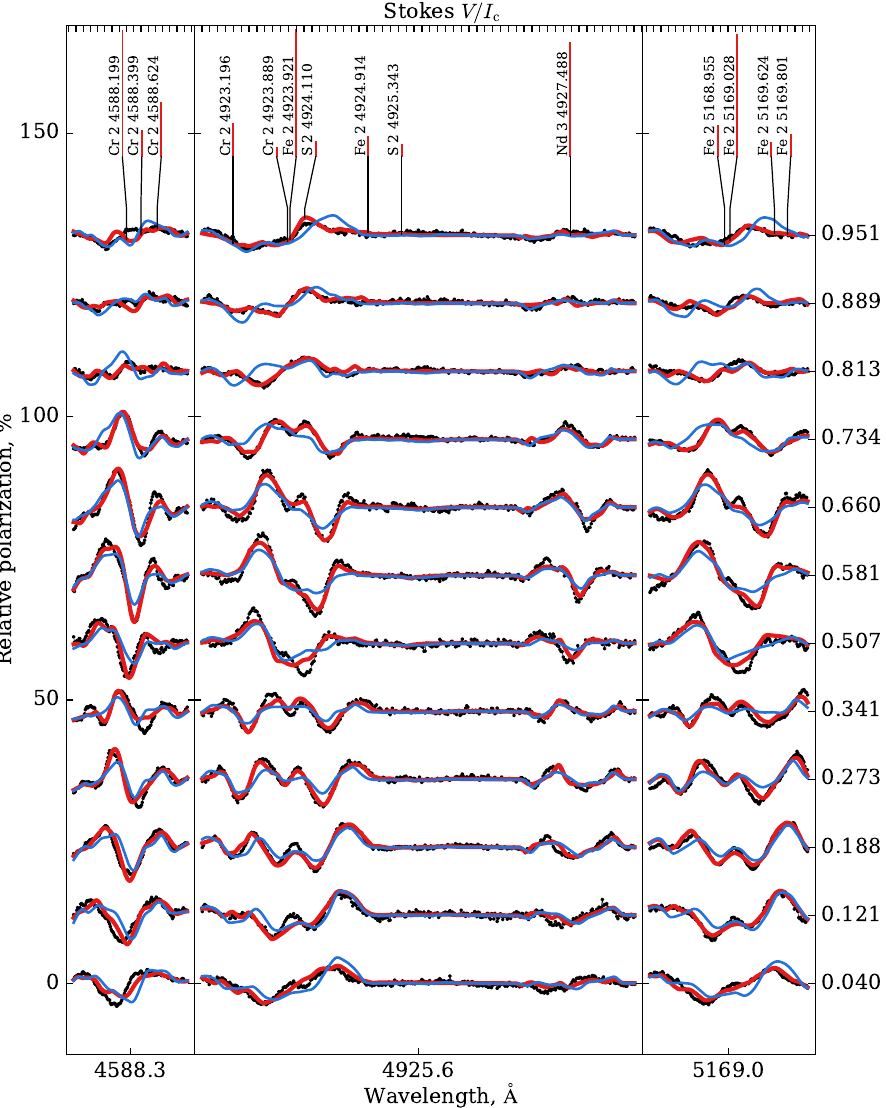}
\vspace*{0.5cm}
        \includegraphics{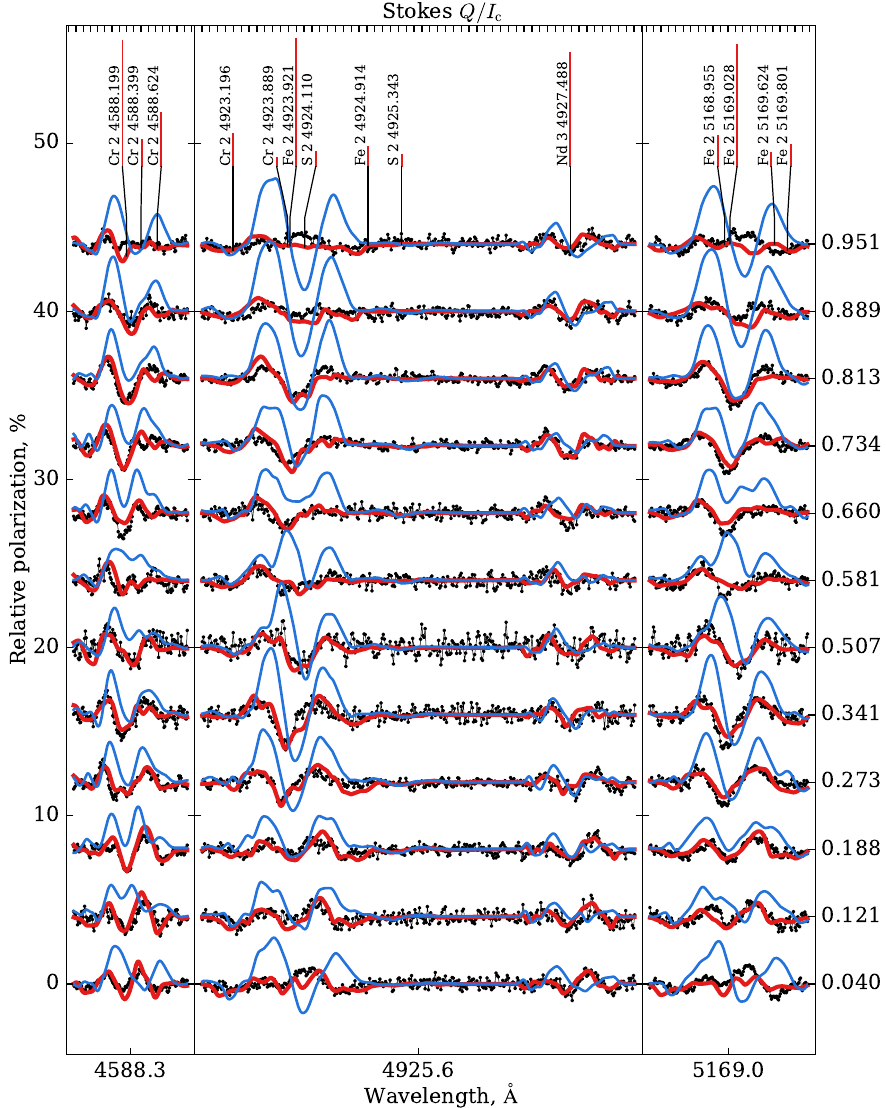}
        \includegraphics{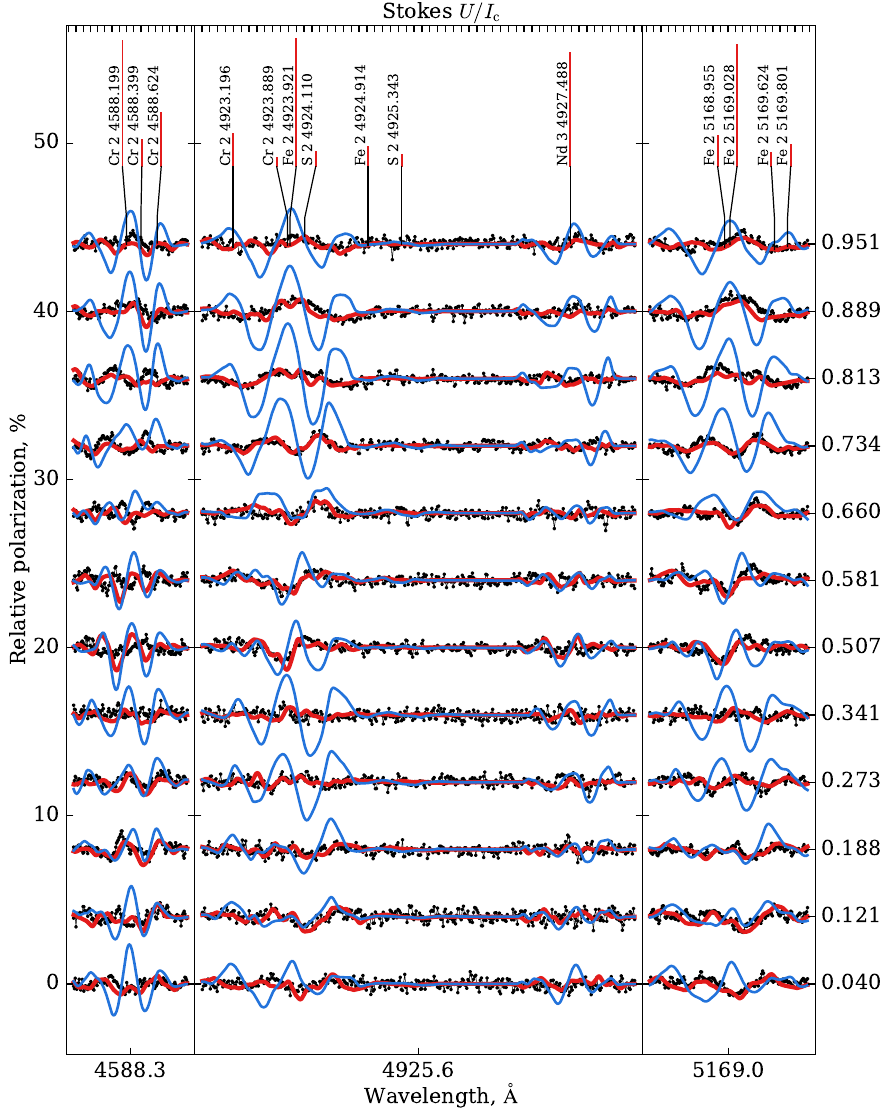}
        \caption{Comparison of the observed (dots connected with black lines) and synthetic Stokes profiles calculated for $\ell\le2$ part of the magnetic field geometry (thin blue lines) and the complete best-fitting MDI solution (thick red lines). The format of this Figure is similar to Fig.~\ref{fig:mdi:StokesIV}.}\label{fig:mdi:comparison}
\end{figure*}

In a further attempt to assess importance of higher-order modes we performed forward calculations considering only the $\ell\le2$ harmonic components from the best-fitting field geometry obtained with the \invers{10} code. Such magnetic field configuration corresponds to a generalised dipole plus quadrupole field geometry. 
Figure~\ref{fig:mdi:comparison} shows a comparison between the line profiles corresponding to $\ell_\mathrm{max}=1$--2 of the harmonic expansion and the solution with $\ell_\mathrm{max}=10$ as used for our definitive MDI magnetic field map described above. 
We only plot a few spectral lines, which are representative of the entire set. 
It is obvious that removing $\ell\ge3$ modes yields a poor fit to observations. In particular, the amplitude of the Stokes $Q$ and $U$ parameters is grossly overestimated, very similar to the effect previously observed for 53~Cam \citep{Bagnulo2001p889,Kochukhov2004p613}.
The description of the data does not improve if we allow the code to adjust the low-order harmonic coefficients or run a new inversion restricting the expansion to $\ell_{\rm max}=2$.
We consider these results as evidence that the more complex field structures corresponding to the $\ell > 2$ harmonic components are justified by the observational data and play an important role in reproducing detailed shapes of the observed Stokes parameter profiles.

\subsection{Chemical abundance maps}

\begin{figure*}
  \centering
  \includegraphics[width=\textwidth]{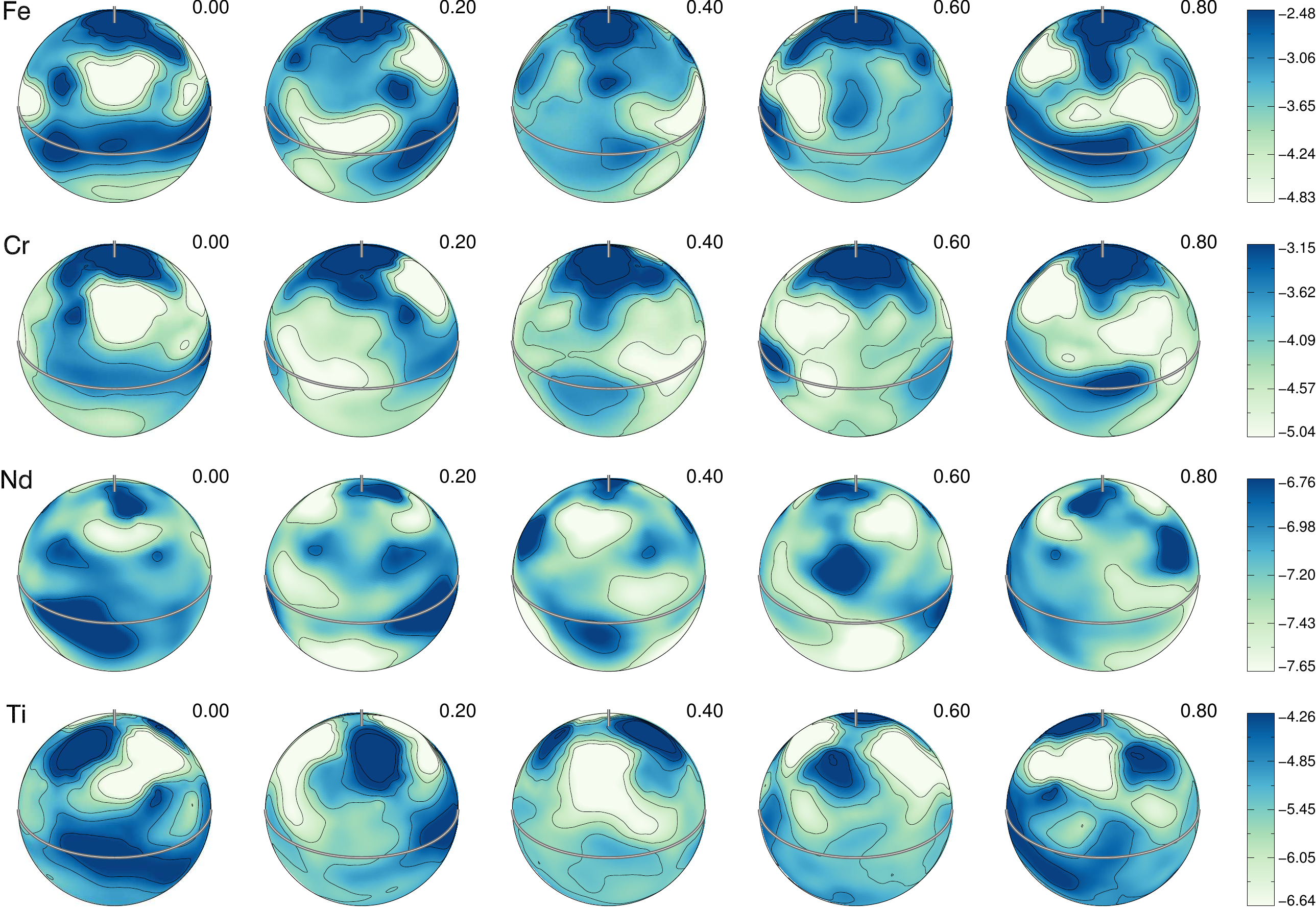}
  \caption{Surface abundance distribution of Fe, Cr, Nd, and Ti. The colour bars on the right indicate the element abundance in $\log (N_X / N_\mathrm{tot})$ units. The contours are plotted with 0.5\,dex increments. The thick line and the vertical bar indicate positions of the rotational equator and the pole, respectively.}\label{fig:mdi:abn}
\end{figure*}

In this study we reconstructed the surface abundance distribution of four chemical elements. The abundance maps of Fe, Cr, and Nd were obtained from fitting the four Stokes parameter spectra, as part of the self-consistent magnetic field inversion. In addition, we derived the abundance distribution of Ti by fitting a set of spectral lines of this chemical element while keeping the magnetic field geometry and abundance maps of other chemical elements fixed. The surface abundance distributions derived for HD\,119419 are illustrated in Fig.~\ref{fig:mdi:abn}. The abundance limits for each map were adopted by considering the 5 and 95 percentiles of each map, so that a small number of extreme abundance value was excluded.

The abundance maps inferred from the MDI inversion show that in general HD\,119419 exhibits very strong surface abundance inhomogeneities of Fe-peak elements. At the same time, the inferred surface abundance distribution of Nd implies a milder surface inhomogeneity. The abundance map of Fe has ranges between $-2.5$ and $-4.8$ in $\log(N_\mathrm{Fe} / N_\mathrm{tot})$ units. Similarly, the inferred abundance map of Cr ranges between $-3.2$ and $-5.0$. It is worth mentioning that both chemical elements exhibit similar features between their abundance maps, corresponding to similar morphological characteristics of spectral line profiles. Nevertheless, we do not find any clear correlation between the abundance distributions of Fe and Cr and the magnetic field geometry of the star. The possible exception is a small overabundance spot at the visible rotational pole, which occurs for both elements. Such features sometimes appear in MDI when the projected rotational velocity of the star has been overestimated. This may lead to the appearance of symmetric overabundance features in the maps, as an inversion code tries to fit the line cores. We performed several experiments with our inversion code, where we reduced the value of the $v_e \sin i$ by 1 and 2\,\kms. However, these inversions did not yield abundance maps that did not feature the aforementioned symmetric overabundance spots, indicating that the polar overabundance spots might not be spurious features of the inversions. We note that similar features have also been observed in other MDI studies, for example, for HD\,24712 \citep{Rusomarov2015p123}.

The surface abundance of Nd has values in the range between $-6.8$ and $-7.7$ in $\log(N_\mathrm{Nd} / N_\mathrm{tot})$ units, which is much smaller than the abundance inhomogeneities seen for Fe and Cr. The surface abundance pattern of Nd does not show any correlation with the abundance maps of Fe and Cr or with the inferred magnetic field geometry.

The surface abundance distribution of Ti was determined from two \ion{Ti}{ii} lines, $\lambda$ 4468.49 and 4571.97~\AA, that had visible polarisation signatures in their Stokes parameters and exhibited significant rotational modulation. All other Ti lines appear to be strongly blended by lines of chemical elements other than Fe, Cr, and Nd. The abundance distribution of Ti was derived by allowing our magnetic inversion code to search for an abundance map of Ti that could reproduce the observed line profiles. In this procedure, the magnetic field configuration and the abundance maps of Fe, Cr, and Nd that were previously derived were not allowed to be modified by the inversion procedure. The resulting abundance map of Ti is presented in Fig.~\ref{fig:mdi:abn} (bottom row). From this Figure it is evident that the abundance distribution of this element to some extent follows the morphology of the Fe and Cr abundance maps. The local Ti abundance has values in the range from $-4.3$ to $-6.6$ in $\log(N_\mathrm{Ti} / N_\mathrm{tot})$ units, and exhibits abundance contrast similar to the Fe and Cr maps.

\begin{figure*}
        \centering
        \includegraphics{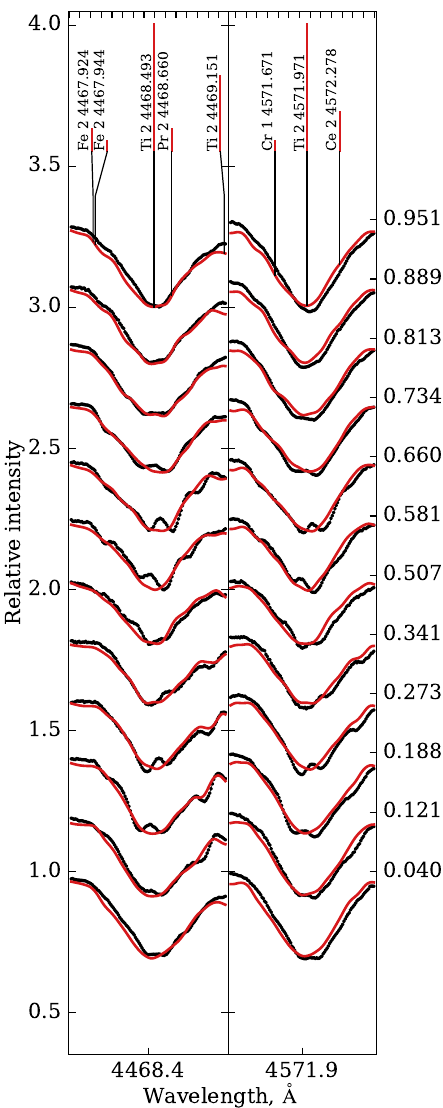}
        \includegraphics{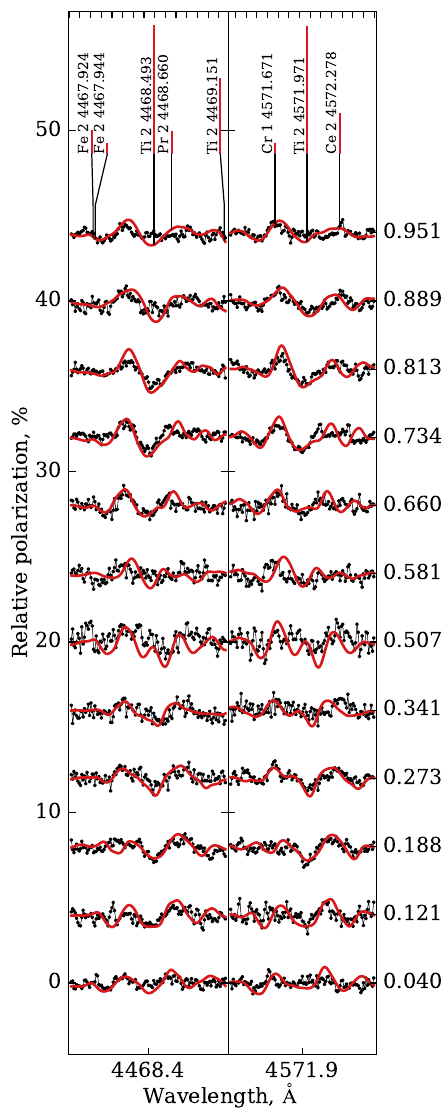}
        \includegraphics{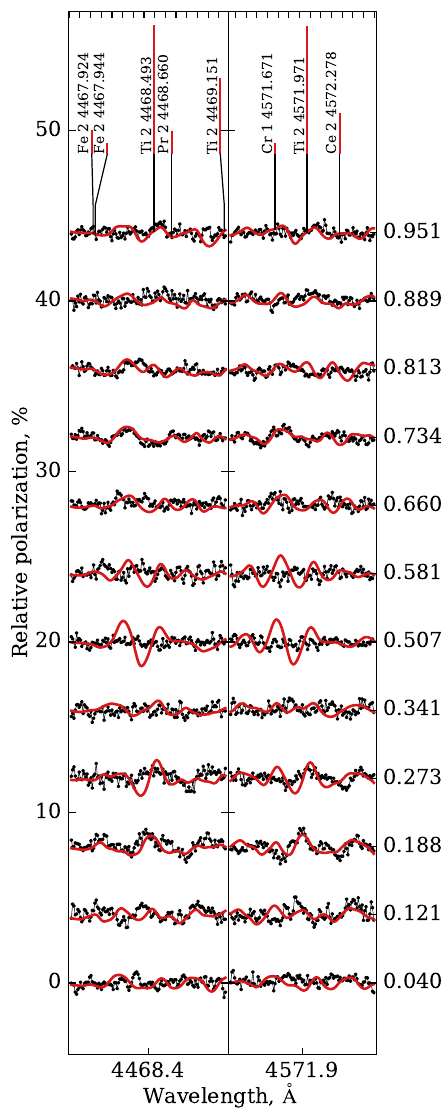}
        \includegraphics{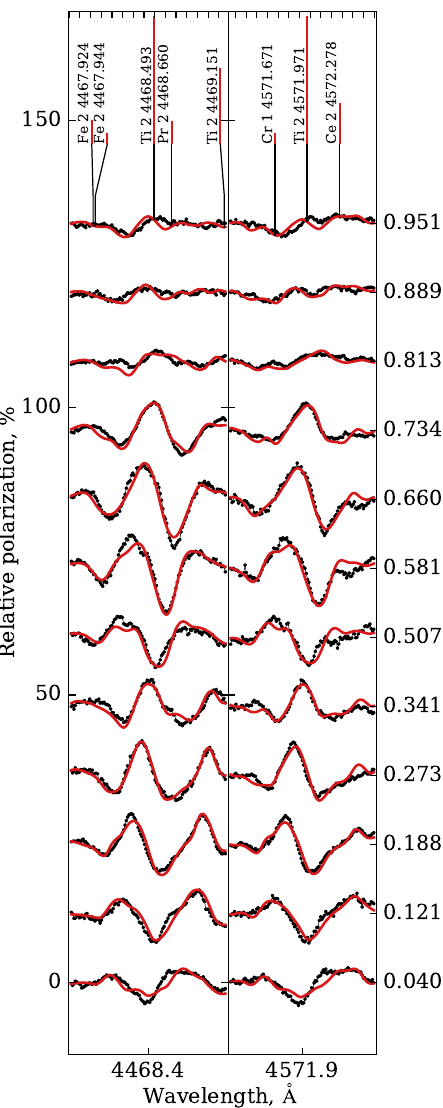}
        \caption{Comparison of the observed (dots connected with black lines) and synthetic (thin red lines) Stokes profiles of Ti lines. The panels show, from left to right, the Stokes~$I$, $Q$, $U$, and $V$ parameter profiles. The format of this Figure is similar to Fig.~\ref{fig:mdi:StokesIV}.}\label{fig:mdi:comparison-Ti}
\end{figure*}

The comparison between the observed and computed profiles of the spectral lines used in the abundance mapping of Ti is presented in Fig.~\ref{fig:mdi:comparison-Ti}. We can see from this Figure that the overall shape of the Stokes $I$ profiles is well reproduced except small details in the line cores. The polarisation profiles are fitted well, even though the magnetic field topology was not allowed to change to improve the fit to the Stokes $QUV$ data. A systematic problem of an overestimation of the Stokes~$QU$ profile amplitude appears for the rotational phases 0.507 and 0.581. However, the S/N of the spectra for these two phases is worse compared to the spectra for other rotational phases, resulting in a solution that produces a systematically worse observed line profile description for these two phases. 

\section{Summary and conclusions}\label{sec:summary}

In this paper we present a model atmosphere analysis as well as a detailed study of the surface magnetic and chemical structure for the magnetic Ap star HD\,119419. This investigation extends our series of four Stokes parameter magnetic mapping studies \citep{Kochukhov2004p613,Kochukhov10p13,Silvester2014p182,Silvester2015p2163,Silvester2017,Rusomarov2015p123,Rusomarov2016p138} to a late-B chemically peculiar star which is a member of an open cluster and therefore has a well-established age.

Based on the high-resolution Stokes $IQUV$ spectra collected at ESO and archival photometric data, we confirm the previously determined effective temperature of HD\,119419, $T_{\rm eff}=11150$~K, and revise the surface gravity to $\log g = 4.3$ by fitting the wings of the hydrogen Balmer lines. We also estimate mean chemical abundances and have taken them into account in calculation of the stellar model atmosphere. The circular polarisation observations of HD\,119419, processed with the least-squares deconvolution technique, allowed us to obtain a new set of high-precision mean longitudinal magnetic field measurements and to determine an improved value of the stellar rotation period, $P_{\rm rot}=2.60059$~d, considering all longitudinal field measurements obtained for this star over the time period of nearly 30 years.

Considering the LSD Stokes $QU$ profiles, we find an interesting discrepancy between the amplitudes of the Fe-peak and REE linear polarisation signatures. This anomaly was confirmed with the analysis of individual lines of these elements. At the same time, we could not measure a statistically significant net linear polarisation from the LSD Stokes $QU$ profiles, likely due to a lack of strong, saturated metal lines in the spectrum of HD\,119419 owing to its relatively high effective temperature.

We then carried out detailed modelling of the surface structure of HD\,119419 by applying the magnetic Doppler imaging method to a set of four Stokes parameter profiles of 14 Fe, Cr, and Nd spectral lines. This magnetic inversion provides detailed information on the stellar magnetic field geometry as well as distributions of the three chemical elements. As a by-product, we also obtained an accurate estimate of the projected rotational velocity, $v_{\rm e}\sin i=25$~\kms, and the two angles specifying the three-dimensional orientation of the stellar rotational axis, $i=60\degr$ and $\Theta=170\degr$.

The largest contribution to the global magnetic field of HD\,119419, in terms of the magnetic field energy, comes from a dipolar component. At the same time, amplitude of the spherical harmonic modes with angular degrees $\ell=2$--5 is non-negligible. These smaller-scale field components are definitely required to properly describe the observed Stokes parameter profiles of HD\,119419. In this respect, HD\,119419 appears to be very similar to 53~Cam and $\alpha^2$~CVn analysed by the first full Stokes vector MDI studies \citep{Kochukhov2004p613,Kochukhov10p13}. In the case of HD\,119419, we have reaffirmed, with additional forward Stokes parameter calculations, that excluding the $\ell\ge3$ harmonic components from the full MDI magnetic field map leads to an inferior description of the observed Stokes $V$ profiles and to an entirely incorrect prediction of the shape and amplitude of the Stokes $Q$ and $U$ spectra. Thus, the octupole and other high-order harmonic terms are clearly essential for fitting the Stokes spectra despite contributing relatively little to the overall magnetic field energy budget.

According to our MDI magnetic field map, the mean surface field strength of HD\,119419 is 17~kG and the local field strength spans from 2 to 24~kG. Except for the phase-averaged mean surface field modulus of $\langle B \rangle=18$~kG, our MDI magnetic field topology bears no resemblance to the field geometry suggested for this star by previous low-order multipolar modelling of the integral magnetic observables \citep{Bagnulo1999p158,Glagolevskii2001,Bagnulo2002p1023}.

Chemical abundance distributions, reconstructed for Fe, Cr, and Nd simultaneously with mapping the field geometry and using a fixed magnetic field geometry for Ti, revealed complex chemical spot maps with varying abundance contrast. The Fe-peak elements exhibit about 2~dex contrast whereas the Nd map is characterised by the local abundance range of approximately 1~dex. We could not establish a clear link to the magnetic field map for any of these chemical spot distributions.

In addition to expanding the still relatively short list of ApBp stars investigated with the four Stokes parameter MDI technique, our study added qualitatively to the research field by presenting the first Stokes $IQUV$ inversion analysis for a cluster Ap star. This is significant because all previous targets of MDI in all four Stokes parameters (HD\,24712, 49\,Cam, 53\,Cam, $\alpha^2$~CVn, CS\,Vir, and HD\,32633) are field stars for which age determination is notoriously unreliable \citep{Landstreet2007}. Therefore, one has to turn attention to cluster Ap stars for meaningful comparison of magnetic inversion results with theoretically predicted evolutionary changes of fossil magnetic field geometries \citep{Braithwaite2006,Duez2010p58}.

A systematic MDI analysis, though likely to be limited to the Stokes $I$ and $V$ spectra, is currently underway for a sample of Ap cluster stars studied by \citet{Landstreet2007} and \citet{Bailey2014}. The first study in that series was recently presented by \citet{Kochukhov2017a} for the young Bp star HD\,133880. Such investigations offer interesting prospects for probing secular evolution of the surface magnetic field topologies and chemical element distributions, as cannot be done with an adequate precision for any sample of field ApBp stars. However, the sample of MDI targets with well-determined ages needs to be expanded to at least ten or so objects with a decent age spread before one can draw any conclusions and compare with theoretical predictions.

\begin{acknowledgements}
This study was supported by the Knut and Alice Wallenberg Foundation, the Swedish Research Council, and the Swedish National Space Board. 
\end{acknowledgements}


\end{document}